\documentclass[a4paper,11pt]{article}
\usepackage[utf8]{inputenc}
\usepackage{mathtools}

\usepackage[T1]{fontenc}
\usepackage{jcappub} 
\usepackage{float}
\usepackage{caption}
\usepackage{subcaption}

\graphicspath{{./}{figures/}}
\usepackage[title]{appendix}
\begin{document}

\newcommand{\etal}{{\it et al.}}
\newcommand{\apjl}{{Astrophys. J. Lett.}}
\newcommand{\apj}{{Astrophys.~J.}}
\newcommand{\apjs}{{ApJS}}
\newcommand{\aap}{{Astron. Astrophys.}}
\newcommand{\aj}{{Astron. J.}}
\newcommand{\mnras}{{Mon.~Not.~Roy.~Astron.~Soc.}}
\newcommand{\na}{{New Astronomy}}
\newcommand{\nat}{{Nature}}
\newcommand{\araa}{{Ann. Rev. of Astronomy \& Astrophysics}}
\newcommand{\jcap}{{JCAP}}
\newcommand{\pasp}{{Publications of the Astronomical Society of the Pacific}}
\newcommand{\prd}{{Phys. Rev. D}}
\newcommand{\prl}{{Phys. Rev. Lett.}}

\newcommand{\mycomment}[1]{}

\title{Thomson scattering: One rate to rule them all}
\author{Kylar L. Greene}
\author{and Francis-Yan Cyr-Racine}
\affiliation{Department of Physics and Astronomy, University of New Mexico \\
210 Yale NE,
Albuquerque, NM 87131, USA}
\emailAdd{kygreene@unm.edu}
\emailAdd{fycr@unm.edu}

\abstract{

The enduring tension between local and distant measurements of $H_0$ remains unresolved.
It was recently pointed out that cosmic microwave background (CMB) and large-scale structure (LSS) observables are invariant under a uniform rescaling of the gravitational free-fall rates of all species present and the Thomson scattering rate between photons and electrons. 
We show that a unique variation of the fine-structure constant $\alpha$ and the electron mass $m_{\rm e}$ can leverage this scaling transformation to reconcile the CMB and LSS data with a broad spectrum of Hubble constant values, encompassing those inferred from local measurements.
Importantly, this study demonstrates that the constraints on the variation of fundamental constants imposed by the specific recombination history are not as stringent as previously assumed.
Our work highlights the critical role of the Thomson scattering rate in the existing Hubble tension and offers a distinct avenue of exploration for particle model builders.
}

\maketitle
\flushbottom

\section{Introduction}

    A well established inconsistency, known as the Hubble tension, is posing a major challenge to the established models of cosmology and particle physics \cite{Planck:2018vyg,aiola:2020,SPT-3G:2021eoc,Riess:2011yx,freedman2012,Riess:2016jrr,CSP:2018rag,Riess:2020fzl,Riess:2021jrx,Krishnan:2021dyb,Krishnan:2021jmh,Luongo:2021nqh}. 
    More specifically, the tension is caused by the divergence of local and distant estimates of cosmological distance scales leading to discrepancies in the observed rate of expansion of the Universe \cite{Efstathiou:2020wxn,Efstathiou:2021ocp,Greene:2021shv}. 
    Local measurements typically employ Type Ia supernovae calibrated with local anchors, while distant scales are inferred from Cosmic Microwave Background (CMB) observations. 
    These distinct approaches calibrate the redshift-distance relation differently, leading to the peculiar observation that locally determined distances often place celestial objects physically closer than their CMB-derived counterparts.
    This intriguing discrepancy between distance measurements does not only present an issue to be resolved, but also a unique opportunity to explore physics beyond the standard models of both cosmology and particle physics.

    The path to resolving the Hubble tension may lie either in modifications to the local or distant Universe.
    Both strategies entail invoking physics beyond the Standard Model, typically by introducing new dynamics within the unseen dark sector.
    Efforts towards a model that locally resolves this discrepancy, whilst maintaining unchanged distances to Type Ia supernovae, are underway, albeit challenging \cite{Mortonson:2009qq,Krishnan:2020vaf,Colgain:2022nlb,Colgain:2022rxy,Malekjani:2023dky,Dhawan:2020xmp,Jassal:2005qc,Caldwell:2003vq,DiValentino:2020naf,Benevento:2020fev,Alestas:2020zol,Keeley:2019esp,Dutta:2018vmq,Alestas:2021luu,Clark:2020miy,Braglia:2020iik,Ballardini:2020iws,Karwal:2021vpk,Nunes:2021zzi,Yang:2018euj,Yang:2018qmz,Yang:2018uae,Yang:2020ope,Yang:2020uga,Yang:2021flj,Yang:2021eud,Yang:2021hxg,Rezazadeh:2022lsf,DiValentino:2016hlg,DiValentino:2017rcr,DiValentino:2019exe,DiValentino:2019ffd,DiValentino:2019jae,DiValentino:2020kha,DiValentino:2020kpf,DiValentino:2020leo,DiValentino:2020vnx,DiValentino:2021rjj,DiValentino:2021zxy}.
    Consequently, our focus shifts towards innovations in the distant Universe \cite{Bernal:2016gxb,Verde:2019ivm,Knox:2019rjx,Paul:2021abc, Schoneberg:2022ggi}.
    This endeavor is undoubtedly challenging due to the tight constraints imposed by detailed observations of the CMB and large scale structure (LSS) \cite{Planck:2018vyg,aiola:2020,SPT-3G:2021eoc}.
    Nonetheless, recent explorations into the Universe's dynamics have uncovered an overlooked symmetry of cosmological observables, known as the Free-Fall, Amplitude, and Thomson (FFAT) scaling \cite{Cyr-Racine:2021oal}.
    This scaling embodies a universal rescaling of the Universe's various length and rate scales, permitting significantly higher values of the Hubble parameter ($H$) while preserving CMB spectra and LSS observations.

    Essential to the FFAT scaling is the rescaling of the Thomson scattering rate in the early Universe.
    The Thomson scattering rate is the dominant process controlling photon diffusion in the pre-recombination plasma and is the second most important rate to understanding the structure of the CMB anisotropies, behind only the background expansion rate.
    The scattering rate is often assumed to be a fixed feature of the cosmological model, but its variation is essential to exploring the overarching tension.
    Indeed, CMB spectra are sensitive to the ratio $\frac{\dot{\kappa}(z)}{H(z)}$, where $\dot{\kappa}(z)$ is the Thomson scattering rate with respect to redshift $z$.
    Any variation to $H(z)$ without equally scaling $\dot{\kappa}(z)$ will change the timing of photon decoupling, the thickness of the last scattering surface, and the suppression of power at high $\ell$ in CMB spectra.
    The observational constraints of the CMB may imply a relationship between the Hubble tension and the Thomson scattering rate.

    While our comprehension of the relationship between the Hubble tension and the FFAT scaling is yet nascent in contrast to other established approaches like early dark energy models \cite{Poulin:2018cxd,Agrawal:2019lmo}, progress is being made exploring this new model space.
    Ref.~\cite{Cyr-Racine:2021oal} effectively scaled $\dot{\kappa}$ in the FFAT scaling direction, generating CMB spectra that aligns with observational data and allows for higher values of $H$. 
    Their methodology, however, necessitated a decrease in the primordial helium abundance, a parameter stringently constrained by Big Bang Nucleosynthesis (BBN).
    Inspired by this line of research and the tantalizing opportunity to explore new physics, we propose a different path: varying the electromagnetic fine-structure constant and the electron mass to achieve the desired scaling of $\dot{\kappa}$. 
    The idea of varying fundamental constants is not new, it was first suggested by Dirac in the 1930s, who proposed that the values of physical constants could be related to the age of the Universe \cite{Dirac:1937aa,Dirac:1938aa}.
    Presently, lab measurements and astrophysical observations restrict possible variations of $\alpha$ and $m_{\rm e}$ up to a redshift of $z\approx7$, assuming the constants vary as a function of a simple power law.
    The fundamental constant variation also affects the recombination history which was previously thought to be well constrained in the $\Lambda$ Cold-Dark-Matter ($\Lambda$CDM) model of cosmology \cite{Planck:2014ylh,Zahn:2002rr,Zaldarriaga:1995gi,Kaplinghat:1998ry,Hannestad:1998xp}.
    In this work, we highlight the protective role of the FFAT scaling against alterations to the recombination history caused by variations in fundamental constants. 
    This relaxation of observational constraints set by the $\Lambda$CDM model on fundamental constant variation widens the scope for model builders to delve into the FFAT scaling direction without disrupting other cosmological observables.
    We demonstrate the efficacy of the FFAT scaling and fundamental constant variation to reconcile CMB, baryon acoustic oscillations (BAO), and local observations of $H_0$, without worsening the fit to any of these observations.
    This constitutes a significant stride towards comprehending the interplay between the Thomson scattering rate and the Hubble tension.

    The structure of this paper is as follows: Section \ref{sec:meth} elucidates the FFAT scaling, which upholds the invariance of cosmological observables under changes to the background expansion rate $H$.
    A specific fundamental constant variation, constructed to ensure near invariance of the recombination history when paired with the FFAT scaling, is detailed in section \ref{sec:fundamental constant variation}.
    Our method of integrating the gravitational effects of the FFAT scaling with the fundamental constant variation into the cosmological code CLASS is delineated in section \ref{sec:toy}.
    Subsequently, section \ref{sec:results} demonstrates how our approach broadens the posterior for $H_0$, accommodating the locally measured value of $H_0 = 73.04 \pm 1.04$ km/s/Mpc \cite{Riess:2021jrx}.
    Finally, in section \ref{sec:conc}, we encapsulate our findings and suggest potential directions for further investigation into resolving cosmological tensions using this technique.
    Note that throughout this paper, we use units in which $c = \hbar = k_{\rm{B}} = 1$.

\section{The FFAT scaling} \label{sec:meth}
    The FFAT scaling, first presented in ref.~\cite{Cyr-Racine:2021oal} and used in ref.~\cite{Ge:2022qws}, provides a powerful tool to understand the origins of cosmological constraints on new physics parameters. 
    The transformation consists of a simple rescaling of rate and length scales by a constant factor $\lambda$ that play key roles in the dynamical evolution of the Universe. 
    Since CMB and LSS predictions are left invariant under the transformation, it also allows the exploration of theoretical models that predict a high value of $H_0$ that are automatically compatible with these observations.
    The set of scaling transformations is given by
        \begin{equation} \label{eq:ffat}
        \begin{aligned}
            \textbf{FF:} \: \sqrt{G\rho_i} & \xrightarrow{\lambda} \lambda \sqrt{G\rho_i}, \\
            \textbf{A:} \: A_{\rm{s}} & \xrightarrow{\lambda} A_{\rm{s}}/\lambda^{n_{\rm{s}}-1}, \\
            \textbf{T:} \: \sigma_{\rm{T}}n_{\rm{e}} & \xrightarrow{\lambda} \lambda \sigma_{\rm{T}}n_{\rm{e}},
        \end{aligned}
        \end{equation}
    where $\xrightarrow\lambda$ indicates the FFAT scaling, $G$ is Newton's gravitational constant, $\rho_i$ is the energy density of the $i^{\rm{th}}$ component, $A_{\rm{s}}$ is the amplitude of the primordial curvature power spectrum, $n_{\rm{s}}$ is the scalar spectral index, $\sigma_{\rm{T}}$ is the Thomson cross-section, and $n_{\rm{e}}$ is the free electron number density.
    We would like to emphasize that $\lambda$ is not a physical parameter in the model, but rather $\lambda$ captures the change in $H$ of a derived model compared to the baseline model, like $\Lambda$CDM.
    A restricted version of this transformation (i.e.~{\bf FF + A} in the above), valid only on very large scales, was initially presented in ref.~\cite{Zahn:2002rr}. 
    By adding the Thomson rate rescaling ({\bf T}), ref.~\cite{Cyr-Racine:2021oal} showed that this transformation leaves CMB and LSS observables invariant over a broad range of scales. 
    This paper focuses on a possible physical implementation of a transformation involving variation of fundamental constants and the \textbf{T} scaling. 
    For clarity, we briefly review each element of the FFAT scaling, emphasizing their role in maintaining the invariance of cosmological observables.

\subsection{Free Fall} \label{subsec:ff}
    The scaling of the gravitational free fall rate (\textbf{FF}), $\sqrt{G\rho_i} \xrightarrow{\lambda} \lambda \sqrt{G\rho_i}$, is possible by scaling the energy density $\rho_i \xrightarrow{\lambda} \lambda^2 \rho_i$\footnote{The FF transformation can also be achieved by scaling the gravitational constant $G$ (as in ref.~\cite{Zahn:2002rr}), perhaps including the FFAT transformation into a theory of modified gravity. 
    }$^,$\footnote{For species with non-constant equations of state, such as massive neutrinos, it's imperative to scale the pressure in the same way as the density.}.
    The energy density of the Universe is directly related to the expansion rate, $H(z)$, as described by the Friedmann equation \cite{Friedman:1922kd}
        \begin{equation} \label{eq:scaledfriedmann}
            H(z) = \sqrt{\frac{8\pi G\rho_i(z)}{3}}.
        \end{equation}
    Therefore, the FFAT scaling enables model builders to increase the value of $H(z)$ such that $H \xrightarrow{\lambda} \lambda H$.
    A model realizing $\lambda \sim 1.08$ would bring the CMB inference of $H_0$ into an agreement with the locally observed value and resolves the Hubble tension without putting other observations into further disagreement if the complete FFAT scaling is realized \cite{Planck:2018vyg,Riess:2021jrx}.
    The \textbf{FF} scaling keeps key cosmological quantities unchanged, including; (1) cosmological angles, (2) the redshift of matter-radiation equality, (3) the free-streaming fraction of the radiation density $\frac{\rho_{\rm{fs}}}{\rho_{\rm{r}}}$, and (4) the matter clustering fraction. 

    \begin{itemize}
          \item \textit{Cosmological Angles --} The FFAT scaling leaves cosmological angles tied to the baryon-photon sound horizon invariant. 
          This includes the angular size of the acoustic sound horizon as measured directly in CMB observation $\theta_{\star}$, as well as the angular size of the baryon acoustic oscillation (BAO) feature imprinted on the distribution of galaxies at lower redshifts. To understand this invariance, consider $\theta_{\star}$, which is given by 
        \begin{equation} \label{eq:thetastar}
            \theta_{\star} = \frac{r_{\star}}{D(z_{\star})},
        \end{equation}
        where $r_{\star}$ is the comoving sound horizon, $D(z_{\star})$ is the comoving angular diameter distance, and a $\star$ references the value at the surface of last scattering.
        $r_{\star}$ and $D(z_{\star})$ are given by
        \begin{equation} \label{eq:rstar}
            r_{\star} = \int_{z_{\star}}^{\infty} \frac{c_{\rm{s}}(z)}{H(z)}\rm{d}z,
        \end{equation}
        \begin{equation}\label{eq:dstar}
            D(z_{\star}) = \int_0^{z_{\star}} \frac{1}{H(z)}\rm{d}z,
        \end{equation}
        where $c_{\rm{s}}$ is the sound speed of the photon-baryon fluid given by $c_{\rm{s}} = 1/\sqrt{3\left(1+\frac{3\rho_{\rm{b}}}{4\rho_{\rm{\gamma}}}\right)}$ \cite{Hu:1996vq,Jedamzik:2020zmd}.
        Since the FFAT scaling leaves ratios of energy densities invariant, 
        the sound speed $c_{\rm{s}}$ is left unchanged.
        Equations \eqref{eq:rstar} and \eqref{eq:dstar} are then both scaled by a factor of $1/\lambda$ (from the $1/H(z)$ factor), leaving the ratio $\theta_{\star} = r_{\star}/D(z_{\star})$ invariant.\footnote{In principle, the integration bounds of equations \eqref{eq:rstar}-\eqref{eq:dstar} are shifted when generally changing the background cosmology. However, as detailed later in section \ref{sec:Xehistory}, the FFAT scaling ensures that the redshift of decoupling is left invariant compared to the $\Lambda$CDM model.}
        Similarly, the FFAT scaling leaves the angular size of the baryon acoustic oscillation (BAO) feature imprinted on the distribution of galaxies at lower redshifts invariant.

      \item \textit{Redshift of Matter-Radiation Equality --}
        The redshift of matter-radiation equality, $z_{\rm{eq}}$, marks the transition from  radiation domination to matter domination in the Universe.
        CMB spectra observations are sensitive to $z_{\rm{eq}}$ via the early integrated Sachs-Wolfe effect (eISW) where photons interact with a time-dependent gravitational potential and introduce additional CMB anisotropies \cite{White:1997vi,Sachs:1967er,Hou:2011ec,Cabass:2015xfa}.
        To keep the CMB temperature spectra invariant under the FFAT scaling, $z_{\rm{eq}}$ must remain invariant.
        $z_{\rm{eq}}$ is given by
        \begin{equation}
            z_{\rm eq} = \frac{\rho_{\rm m}^0}{\rho_{\rm r}^0}-1
        \end{equation}
        where $\rho_{\rm m}^0$ and $\rho_{\rm r}^0$ are the present-day energy densities of matter and radiation.
        Similarly, the FFAT scaling leaves energy density ratios invariant, so the eISW effect is left unchanged under the FFAT scaling, which in turn leaves the low $\ell$ $C_{\ell}$'s invariant.
        
      \item \textit{Free-Streaming Radiation Ratio --}
          Free-streaming particles that propagate faster than the sound speed of the pre-recombination plasma contribute to the decay of gravitational potentials which drive acoustic oscillations \cite{Bashinsky:2003tk,Follin:2015hya,Baumann:2015rya}.
          The free-streaming to radiation fraction is given by
          \begin{equation}
              f_{\rm fs} = \frac{\rho_{\rm fs}}{\rho_{\rm r}}
          \end{equation}
          where $\rho_{\rm fs}$ is the energy density of free-streaming radiation.
          Including additional free streaming radiation pulls on the wavefront of the slower propagating plasma, smoothing the perturbation \cite{Cyr-Racine:2013jua}.
          This damping would have noticeable effects on the CMB by introducing a phase and amplitude shift into the peaks of the temperature spectra.
          Again, the FFAT transformation uniformly scales all energy densities such that $\rho_{\rm fs}$ and $\rho_{\rm r}$ are evenly scaled, and $f_{\rm fs}$ is left invariant.
      \item \textit{Matter Clustering Fraction --}
        The growth of matter structure at early times is dictated in part by the fraction of the total matter density that is pressure-supported \cite{Addison:2013haa}. Indeed, baryons cannot contribute to density fluctuations until the end of the baryon drag epoch, leaving only the dark matter to establish gravitational potential wells prior to this time. These potentials impact the CMB through the Sachs-Wolfe effect and set up the initial conditions for what would eventually become the large-scale structure of the Universe. Subtle changes to these potential is tightly bounded by the data and is a major source of constraints on physics beyond $\Lambda$CDM \cite{Ge:2022qws}.
        Since the {\bf FF} transformation equally scales the pressure-supported and non-supported matter content of the Universe, it leaves these gravitational potentials unchanged. Once coupled to the {\bf A} and {\bf T} scalings (see below), the {\bf FF} transformation thus keeps observables related to matter clustering such as $\sigma_8$, the matter power spectrum, and lensing of the CMB invariant.

    \end{itemize}
    
\subsection{Amplitude} \label{subsec:amp}
 
    The Boltzmann equations that describe the evolution of photon, baryon, neutrino, and dark matter perturbations depend on the Fourier wavenumber $k$. This introduces another length scale into the problem (beyond the Hubble and Thomson rates; see next section for the latter) which must be appropriately scaled (i.e.~$k \xrightarrow {\lambda} \lambda k$) to leave the equations of motion invariant.
    However, as shown in references \cite{Zahn:2002rr,Ge:2022qws}, one could bypass this scaling by noting that the solution to the Boltzmann equations $\Tilde{\Phi}$ under the scaling of the Hubble and Thomson rates can be written in terms of the original solution $\Phi$ (i.e.~in the absence of scaling) but with a different wavenumber $k^{\prime} = k/\lambda$,
    \begin{equation} \label{eq:pertur}
        \Tilde{\Phi}(k,a,\lambda) = \Phi(k^{\prime},a,\lambda=1).
    \end{equation}
    Here, $\Phi,\Tilde{\Phi}$ stand for any perturbation variable entering the Boltzmann equations. The above identity is important as it allows us to show that the impact of simultaneously scaling the Hubble and Thomson rates on the CMB and large-scale structure can be entirely absorbed into a modification of the primordial amplitude of scalar fluctuations, as we now show.
    Taking the CMB temperature power spectra as an example, we can write its rescaled version as
    \begin{equation} \label{eq:cmbtt}
        C_{\ell}^{\rm TT}(\lambda) = \int \frac{dk}{k} P(k) |\tilde{\Delta}_{\rm T\ell}(k,\lambda)|^2,
    \end{equation}
    where $P(k)$ is the primordial spectrum of curvature  fluctuations and $\tilde{\Delta}_{\rm T\ell}(k,\lambda)$ is the photon transfer function, which only depends on $\Tilde{\Phi}$ \cite{Lifshitz:1945du,Peebles:1970ag,Bond:1984fp,Bond:1987ub,Vittorio:1984aaz,Wilson:1981yi,Gouda:1992in,White:1995qm,Scott:1995uj,Hu:1995fqa,Bond:1983hb,Ma:1995ey,Dodelson:1995es, Zaldarriaga:1995gi}.\footnote{As noted earlier, the specific scaling of the Thomson scattering rate preserves the visibility function relative to the $\Lambda$CDM model. This ensures a high degree of similarity between the photon transfer functions for both models.} This means that the transfer function obeys the identity $\tilde{\Delta}_{\rm T\ell}(k,\lambda) = \Delta_{\rm T\ell}(k/\lambda,\lambda=1)$.
    We take $P(k$) to have the form
    \begin{equation}
        P(k) = A_{\rm{s}} \left(\frac{k}{k_{\rm{p}}}\right)^{n_{\rm{s}}-1}, 
    \end{equation}
    where $A_{\rm s}$ is the power spectrum amplitude at some pivot wavenumber $k_{\rm{p}}$, and $n_{\rm{s}}$ is the scalar spectral index \cite{Seljak:1996is}. Using the scaling of the photon transfer function, it is straightforward to show that the CMB temperature spectrum is left unchanged by the {\bf FF} transformation if $A_{\rm s}$ is simultaneously scaled by $A_{\rm s} \xrightarrow{\lambda} \frac{A_{\rm s}}{\lambda^{n_{\rm{s}}-1}}$ such that
    \begin{equation}
    \begin{split}
        C_{\ell}^{\rm TT}(\lambda) &\to\int \frac{dk}{k} \frac{A_{\rm{s}}}{\lambda^{n_{\rm s}-1}}\left(\frac{ k}{ k_{\rm{p}}}\right)^{n_{\rm{s}}-1} |\Delta_{\rm T\ell}(k/\lambda,\lambda = 1)|^2 \\
        &= \int \frac{dk^{\prime}}{k^{\prime}} \frac{A_{\rm{s}}}{\lambda^{n_{\rm s}-1}} \left(\frac{\lambda k^{\prime}}{k_{\rm{p}}}\right)^{n_{\rm{s}}-1} |\Delta_{\rm T\ell}(k^{\prime},\lambda = 1)|^2 \\
        &= C_{\ell}^{\rm TT}(\lambda=1).
    \end{split}
    \end{equation}
  A similar argument applies to the polarization and cross temperature-polarization power spectra, as well as to several large-scale structure observables including cosmic shear power spectra.   
  
\subsection{Thomson} \label{sec:thomson}
    After the Hubble expansion rate, the Thomson scattering rate is the next most important rate affecting the structure of the CMB and LSS as it governs interactions between photons and free electrons before and during the recombination epoch.
    It is given by
    \begin{equation}\label{eq:scatteringrate}
        \dot{\kappa} = an_{\rm{e}} \sigma_{\rm{T}},
    \end{equation}
    where $\sigma_{\rm T}$ is the Thomson cross section given by
    \begin{equation}
    \sigma_{\text{T}} = \frac{8\pi}{3} \left(\frac{\alpha}{m_{\rm e}}\right)^2.
    \end{equation}
    $\dot{\kappa}$ controls photon diffusion, and thus the damping of small-scale CMB anisotropies \cite{Silk:1967kq} and polarization generation \cite{Kamionkowski:1996ks, Kamionkowski:1997av}.
    The overall level of photon diffusion is determined by the ratio
    \begin{equation}\label{eq:ratio}
        \frac{\dot{\kappa}(z)}{H(z)}, 
    \end{equation}  
    which must be kept fixed to preserve the shape of CMB temperature and polarization spectra. 
    To preserve equation \eqref{eq:ratio} under a $H  \xrightarrow{\lambda} \lambda H$ scaling , $\dot{\kappa}$ must also be scaled as $\dot{\kappa} \xrightarrow{\lambda} \lambda \dot{\kappa}$. In the familiar language of the CMB sound horizon and Silk damping scales \cite{Aylor:2018drw}, simultaneously scaling both $H(z)$ and $\dot{\kappa}(z)$ automatically leaves both scales invariant. Conversely, any cosmological model that does not preserve the $\dot{\kappa}(z)/H(z)$ ratio lead to a differential change to the CMB sound and damping scales, which typically limits the capability of these models to address the Hubble tension. 
    It is interesting to highlight that, although the scattering rates of photons to baryons differ greatly from those of baryons to photons, attributable to their respective number densities differing by order $10^9$, both species undergo equivalent density scaling. This ensures that the photon to baryon energy density ratio remains consistent with $\Lambda$CDM, thus leaving both scattering rates remain invariant.
    This highlights the key role that Thomson scattering plays in our understanding of cosmology.
    
\subsection{Cosmological anchors: Breaking the scaling invariance}\label{sec:break}
In our Universe, the FFAT scaling invariance is broken by key observations that essentially anchor the transformation, and thus underpin our entire knowledge of cosmology.  
Here we list some of these cosmological anchors, ordered roughly from most impactful to least impactful.

\begin{itemize}

\item \textit{COBE-FIRAS --}
    The CMB spectral energy distribution measurements by COBE-FIRAS indicate a nearly perfect black-body with a temperature of $T_0 = 2.726 \pm 0.002$ K today, placing strict constraints on the photon energy density of the Universe \cite{Fixsen:1996nj,Fixsen:2009ug}.
    This introduces a significant anchoring effect for the \textbf{FF} scaling of the FFAT symmetry, which requires a uniform scaling of all energy densities, including the photon energy density. Getting around this COBE-FIRAS anchor is the most important model-building challenge necessary to fully exploit the FFAT transformation as a means to address the Hubble tension.

\item \textit{Big Bang Nucleosynthesis --}
    BBN depends on the background temperature and expansion rate as they affect the neutron-to-proton ratio at the onset of nucleosynthesis. 
    Changing these factors affects the creation of light elements as the neutron abundance will be changed. 
    The measured primordial helium and deuterium abundances, $Y_{\rm{P}} = 0.2449 \pm 0.0040$ and $D/H = 2.53 \pm 0.15 \times 10^{-5}$ respectively, tightly constrain the Hubble rate during BBN \cite{Aver:2015iza,Pitrou:2018cgg,Consiglio:2017pot,SimonsObservatory:2018koc,Aver:2021rwi,Fields:2019pfx}, and hence anchor the transformation $H \xrightarrow{\lambda} \lambda H$. A way around this BBN constraint is to assume that whatever physics exploits the FFAT scaling to yield a large Hubble constant today was not active at the time of nucleosynthesis. Unless otherwise mentioned, we will work under this assumption in what follows. 
    
\item \textit{Recombination --}
    The ionization history of the Universe near the CMB epoch depends sensitively on the ratio of the hydrogen recombination rate to the Hubble expansion rate. Changing the latter without a corresponding rescaling of the former would thus modify the recombination history, which would in turn change the CMB's last scattering surface. However, since Thomson scattering and hydrogen recombination are similar electromagnetic processes, they have nearly identical parametric dependence, with both rates proportional to $\sigma_{\rm T}n_{\rm e}$ at leading order (i.e.~when neglecting the details of hydrogen atomic physics). This means that the FFAT transformation leaves the recombination history of the Universe \emph{approximately} invariant. It does not, however, leave it \emph{exactly} invariant as other subleading atomic rates also enter the detailed evolution of the hydrogen ionization history, most notably the forbidden $2s\to1s$ transition rate. These residual changes to the recombination history under the FFAT scaling were studied in refs.~\cite{Cyr-Racine:2021oal,Ge:2022qws}, where it was shown that they only become important for relatively large values of $\lambda \gtrsim 1.1$.
\end{itemize}
    
\subsection{Avoiding symmetry-breaking with a Mirror World Dark Sector} \label{subsec:MWDS}
    We now present a phenomenological model which can exploit the FFAT scaling while avoiding the constraints from the prior section. 
    The strictest constraint is imposed by COBE-FIRAS.
    The FFAT scaling requires a uniform scaling across all energy densities, including those of photons.
    However, COBE-FIRAS only detects the \textit{visible} photon energy densities.
    One possible avenue to circumvent the COBE-FIRAS constraint while satisfying the FFAT scaling is to introduce an energy component within the dark sector that behaves as radiation.
    The required dark radiation \cite{Ackerman:2008kmp,Berger:2016vxi} can be included into the dark sector utilizing a ``mirror world dark sector'' (MWDS) containing a copy of photons, baryons, and neutrinos \cite{Chacko:2005un,Chacko:2005vw,Barbieri:2005ri,Craig:2013fga,GarciaGarcia:2015fol,Craig:2016lyx,Farina:2015uea,Prilepina:2016rlq,Barbieri:2016zxn,Craig:2015pha,Chacko:2016hvu,Csaki:2017spo,Elor:2018xku,Hochberg:2018vdo,Francis:2018xjd,Harigaya:2019shz,Ibe:2019ena,Dunsky:2019upk,Csaki:2019qgb,Koren:2019iuv,Terning:2019hgj,Johns:2020rtp,Roux:2020wkp,Ritter:2021hgu,Curtin:2021alk,Curtin:2021spx,Bansal:2022qbi,Foot:2007iy}.
    A MWDS presents a compelling avenue for moving along the FFAT scaling direction, particularly within the realm of dark sector models containing an explicit $Z_2$ symmetry, as exemplified by twin Higgs models \cite{Chacko:2005pe, Bansal:2021dfh}. 
    In this setup, the dark sector mirrors the Standard Model in terms of particle content and gauge interactions, but can have different masses for the corresponding particles. When combined with an Affleck-Dine condensate mechanism which populates both sectors, it inherently circumvents numerous fine-tuning challenges and minimizes the total number of additional parameters required \cite{Blinov:2021mdk}. This includes fine-tuning concerns regarding the timing of the dark sector recombination compared to the visible sector \cite{Roux:2020wkp}. Indeed the approach of ref.~\cite{Blinov:2021mdk}  naturally leads to the ratio of binding energy to background temperature to be similar in both sectors. This preserves the radiation free-streaming fraction evolution, which must be unchanged to stay along the FFAT scaling direction.

    Such a model also naturally explains the additional baryon density requirement.
    We can introduce the required baryon density as atomic dark matter, thus satisfying the baryon scaling requirement and preserving the effective (visible + dark) baryon-photon ratio \cite{Kaplan:2009de, Kaplan:2011yj, Foot:2003jt,Farina:2016ndq, Cyr-Racine:2012tfp, Cyr-Racine:2013fsa}.
    This form of atomic dark matter constitutes a fractional percentage of the total dark matter budget and consists of two fundamental fermions charged under a $U(1)$ symmetry, analogous to the visible proton and electron \cite{Feng:2009mn,Agrawal:2016quu}.
    Lastly, the additional neutrinos ensure the  ratio of free-streaming to coupled radiation energy density remains invariant.

    The next constraint originates from BBN. 
    Present knowledge places limits on fluctuations in $H(z)$ during BBN, considering the formation of light elements is intricately linked to the expansion rate.
    Consequently, we work under the assumption that the mechanism responsible for reheating the dark sector must activate post-BBN, see e.g.~refs.~\cite{Holst:2023hff,Blinov:2021mdk,Allahverdi:2020bys, Aloni:2021eaq,Aloni:2023tff,Joseph:2022jsf,Buen-Abad:2022kgf,Buen-Abad:2023uva}.
    We thus set $Y_{\rm{P}} = 0.2449$ in our model to avoid possible symmetry-breaking effects arising from BBN. Similarly, we assume that the physics responsible for changes in fundamental constants is not present during BBN.

    The final and weakest constraint comes from recombination.
    Recall that along the FFAT scaling direction, recombination is only slightly broken.
    We choose to leave this symmetry breaking intact and in section \ref{sec:toy}, explore its impact numerically. 
    However, implementing the MWDS with FFAT scaling and fundamental constant variation in existing cosmological code poses computational challenges.   
    Consequently, in subsection \ref{subsec:mimic}, we propose an alternative methodology to mimic the MWDS with FFAT scaling into the CLASS code \cite{Lesgourgues:2011re,Blas:2011rf,Lesgourgues:2011rg,Lesgourgues:2011rh}. 
    This approach successfully reproduces the leading-order gravitational effects that the MWDS exerts on the observable Universe, while keeping the recombination history and cosmological observables intact.
    A later publication will present the full MWDS with FFAT scaling implementation, and probe the nature of the more complex dark sector.

\section{Fundamental constant variation} \label{sec:fundamental constant variation}

    To implement the FFAT scaling, one must choose how to scale the Thomson scattering rate to achieve the required phenomenology. 
    Recent studies \cite{Cyr-Racine:2021oal,Ge:2022qws} have opted to increase $\dot{\kappa}$ by scaling the helium fraction $Y_{\rm{P}}$, while keeping the baryon density constant. 
    The scaling $(1 - Y_{\rm{P}}) \xrightarrow{\lambda} \lambda (1 - Y_{\rm{P}})$ effectively raises $n_{\rm e}$ and thus $\dot{\kappa}$ by a factor of $\lambda$, preserving the invariance of the ratio in equation \eqref{eq:ratio}. 
    These studies confirm that the FFAT scaling allows for increased values of $H(z)$ while keeping other CMB and LSS observables in agreement. 
    It however requires a value of $Y_{\rm{P}}$ that is strong tension \cite{Cyr-Racine:2021oal} with direct measurements of the helium abundance \cite{Aver:2015iza,Pitrou:2018cgg,Consiglio:2017pot,SimonsObservatory:2018koc,Aver:2021rwi,Fields:2019pfx}.

    In this study, we postulate that the Thomson cross-section in the early Universe diverged slightly from its current value, rather than change the predicted helium abundance. 
    This deviation suggests that the fundamental constants appearing in the Thomson cross section, specifically the fine-structure constant $\alpha$ and the electron mass $m_{\rm e}$, had subtly different values around the time of decoupling compared to their present-day measurements.
    %While fundamental constant variation has been explored previously in different contexts (see e.g.~refs.~\cite{Webb:1998cq,Murphy:2000pz,Sandvik:2001rv,Uzan:2002vq,Webb:2010hc,Srianand:2004mq,Chand:2004ct,Mota:2003tc,King:2012id,Fischer:2004jt,Dixit:1987at,Mota:2003tm,Coc:2006sx,Huntemann:2014dya,Cowie:1995sz,Petrov:2005pu,Sekiguchi:2020teg,Hart:2017ndk,Hart:2019dxi,Reinhold:2006zn,Calmet:2002ja,Uzan:2004qr,Fritzsch:2016ewd,Flambaum:2007my,Berengut:2010yu,Calmet:2006sc,Chiba:2006xx,Karshenboim:2003jc,Li:2021dvt, Bekenstein:2002wz}), prior works did not consider the invariance of certain cosmological observables protected by the FFAT scaling, thus deriving overly strict constraints on the possible variation of $\alpha$ and $m_{\rm e}$ \cite{Planck:2014ylh,Zahn:2002rr,Zaldarriaga:1995gi,Kaplinghat:1998ry,Hannestad:1998xp}. 
    The following sections elucidate the nature and magnitude of the fundamental constant variations required to realize the desired Thomson rate scaling, while minimally altering the intricacies of hydrogen atomic physics.
    
   \subsection{Thomson cross section rescaling}
    
    When scaling the Thomson cross section, care must be taken to leave the atomic physics of hydrogen nearly invariant. 
    The most important quantity to leave unchanged is the hydrogen binding energy $B_{\rm{H}} \propto m_{\rm{e}}\alpha^2$, which controls the onset of hydrogen recombination and, ultimately, the redshift of photon decoupling.
    In order to meet the condition $\sigma_{\rm{T}}\xrightarrow{\lambda} \lambda\sigma_{\rm{T}}$ without disturbing $B_{\rm{H}}$, a specific variation is necessary as outlined by Ref.~\cite{Ge:2022qws},
    \begin{equation}\label{eq:truescale}
        m_{\rm{e}} \xrightarrow{\lambda} \lambda^{-1/3} m_{\rm{e}} \And \alpha \xrightarrow{\lambda} \lambda^{1/6} \alpha.
    \end{equation}
    Addressing the Hubble tension using the FFAT scaling combined with fundamental constant variation necessitates a value of $\lambda \approx 1.088$ (corresponding to $H_0$ = 73.3 km/s/Mpc) \cite{Riess:2021jrx}. 
    This corresponds to a modest, yet significant, variation of 2.76\% in $m_{\rm e}$ and 1.41\% in $\alpha$ from their respectively measured local values. 
    While these changes might appear small, these early-Universe variations are magnitudes greater than those permissible in the low-redshift Universe (see section \ref{subsec:fcvconstraint}). 
    Also, conventional understanding of fundamental constant variation during the recombination epoch suggests that the Universe's recombination history places constraints on possible variations, a topic that will be revisited in the following subsection.
    The implementation of the FFAT scaling with fundamental constant variations \eqref{eq:truescale} safeguards the crucial ratio depicted in equation \eqref{eq:ratio}, but introduces minor alterations to hydrogen recombination given the different parametric dependence of other atomic rates on $\alpha$ and $m_{\rm e}$ which we discuss in section \ref{subsec:recomb}.
     %We emphasize that any cosmological model that increases $H$ must account for the change in the ratio given by equation \eqref{eq:ratio} to maintain agreement with observables.
    For brevity, the FFAT transformation with fundamental constant variation given by equation \eqref{eq:truescale} will be referred to below as the true FFAT (tFFAT) scaling.

    \subsection{Fundamental constant variation constraints}
    \label{subsec:fcvconstraint}
    A variety of methods, including atomic clock comparisons at $z$ = 0 \cite{Rosenband2008aaa,Cingoz2008aaa,Peik2006aaa,Godun:2014naa}, the Oklo phenomenon (a naturally occurring fission reaction) at $z \approx 0.14$ \cite{Kuroda:1956aa,Shlyakhter:1976aa,Gould:2006qxs}, assessments of radioactive elements derived from meteorite dating at $z \approx 0.45$ \cite{Olive2002:aa}, observations of the Sunyaev-Zeldovich effect at $z \approx 0.5$ \cite{Gall:2017gva}, and inspection of quasar spectra at $z \approx 7$ \cite{Wilczynska:2020rxx,Vernet:2011fy}, offer an extensive set of measurements to constrain the time variation of $\alpha$ and $m_{\rm e}$.
    This is not a complete list of methodologies, and there are several reviews which provide a more complete discussion \cite{Uzan:2010pm, Flambaum:2007aa}.
    The collection of these methods finds that there is nearly no time variation allowed for the constants in the redshift range $0 \leq z \leq 7$.
    However, they do not place bounds on what the allowed level of variation is at the time of photon decoupling, $z \sim 1100$.

    Prior understanding was that the CMB imposed constraints on the permissible variation of constants, due to the variation's inability to keep the recombination history unchanged from the $\Lambda$CDM model.
    References \cite{Planck:2014ylh,Ivanov:2020mfr} determined that the distance to the surface of last scattering, the overall level of diffusion damping \cite{Silk:1967kq}, and the early integrated Sachs-Wolfe effect \cite{Hu1995:aa} are all related and will all be altered under fundamental constant variation.
    The tFFAT scaling preserves these quantities, as discussed in subsection \ref{subsec:ff}, thereby avoiding the constraints set by the CMB on fundamental constant variation.
    This scaling also mostly maintains the integrity of hydrogen recombination, as we will discuss in the next section.
    
    Delving further into redshift space allows us to constrain the permitted variation of constants during BBN at $z \sim 10^9$.
    Alterations in the values of these constants could potentially disturb the predicted abundances of light elements, an outcome not safeguarded against by the FFAT scaling.
    Perturbations to $\alpha$ and $m_{\rm e}$ could possibly modify the lifetime of neutrons and the mass differential between protons and neutrons \cite{Clara:2020efx}. 
    Such changes would adjust the neutron-to-proton ratio, thereby introducing potential deviations in the predicted abundances of light elements.
    Accurately calculating these effects requires intricate QCD and QED lattice calculations \cite{Borsanyi:2014jba}. 
    Therefore, it is of interest to investigate the impact of fundamental constant variation on BBN abundances to constrain the allowed variation of $\alpha$ and $m_{\rm e}$ when combined with the broader FFAT symmetry (particularly the expansion rate scaling) to understand this potential symmetry-breaking effect, but this is beyond the scope of this paper.

    To avoid these constraints set on the variation of fundamental constants, we make the following conservative assumptions; the mechanism which varies the constants must become active after BBN, the time-scale in which the constants vary must be greater than the timescale of recombination, and the constants must be at their stable, locally measured values by $z \sim 7$.

    \subsection{Recombination with fundamental constant variation} \label{subsec:recomb}
    To track how the tFFAT scaling influences the recombination history, we compute the free electron fraction $X_{\rm e}$ and visibility function, limiting our discussion to hydrogen recombination as the visibility function has little support from the specific details of helium recombination.
    In our analysis, we consider quantities not protected by the tFFAT scaling and assume a simplified description of recombination for clarity.
    A more detailed, multi-level approach like that used by the cosmological code HyRec \cite{Lee:2020obi,Ali-Haimoud:2010hou} reveals similar levels of symmetry breaking in appendix \ref{app:4level}.
    
    The Saha approximation (eq.~\ref{eq:saha}) is employed to describe systems in equilibrium and estimate the redshift of the onset of recombination:
    \begin{equation}
        \label{eq:saha}
        \frac{X_{\rm e}^2}{1 - X_{\rm e}} \approx \frac{1}{n_{\rm b}}\left(\frac{m_{\rm{e}} T}{2\pi}\right)^{3/2}e^{-B_{\rm{H}}/T} \equiv s
    \end{equation}
    where $X_{\rm e}$ is the free electron fraction, $n_{\rm b}$ is the visible baryon number density\footnote{In the tFFAT scaling, the visible baryon density is consistent with $\Lambda$CDM, being anchored by BBN predictions. It is up to model builders to introduce additional baryon density without putting BBN constraints into tension. A plausible approach is to invoke a more complex dark sector like that described in section \ref{subsec:MWDS} in which a fractional amount of dark matter is baryonic.}, and $T$ is the background temperature.
    However, it is not invariant under the tFFAT scaling of $m_{\rm e}$, leading to a symmetry-breaking effect of $s \xrightarrow{\lambda} \lambda^{-1/2} s$ .  
    While this effect has negligible influence on the overall $X_{\rm e}$ history, it does shift the equilibrium value of $X_{\rm e}$.
    However, $X_{\rm e}$ falls out of equilibrium at the beginning of recombination and we must transition to a more comprehensive Boltzmann treatment. 
    This transition necessitates the use of the Peebles equation \cite{Peebles:1968ja}, which describes the recombination history by evaluating the rate of change of $X_{\rm e}$ with respect to the cosmological scale factor $a$ and is given by
    \begin{equation}\label{eq:Xehistory}
        \frac{{\rm d}X_{\rm{e}}}{{\rm d}a} = \frac{C\alpha^{(2)}}{aH}\left(\left(\frac{m_{\rm{e}} T}{2\pi}\right)^{3/2} e^{-B_{\rm{H}}/T}(1-X_{\rm{e}}) - n_{\rm{b}}X_{\rm{e}}^2\right).
    \end{equation}
    Here, $\alpha^{(2)}$ is the case B recombination rate given approximately by \cite{Hummer:1987hn}
    \begin{equation}\label{eq:alpha2}
        \alpha^{(2)} \approx \frac{\alpha^2}{m_{\rm{e}}^2}\left(\frac{B_{\rm{H}}}{T}\right)^{1/2}\text{ln}\left(\frac{B_{\rm{H}}}{T}\right).
    \end{equation}
    
    In Equation \eqref{eq:Xehistory}, two factors are not protected by the tFFAT scaling: $\left(\frac{m_{\rm{e}} T}{2\pi}\right)^{3/2}e^{-B_{\rm{H}}/T}$ and $\frac{C\alpha^{(2)}}{aH}$.
    The first term mirrors the symmetry-breaking term in the Saha equation, introducing a factor of $\lambda^{-1/2}$. 
    The second term demands a more thorough examination.
    Specifically, $\alpha^{(2)}$ is not invariant under the tFFAT scaling, where the $\frac{\alpha^2}{m_{\rm e}^2}$ term introduces a symmetry breaking such that $\alpha^{(2)} \xrightarrow{\lambda} \lambda \alpha^{(2)}$.
    Fortunately, this cancels out the factor of $\lambda$ introduced in the denominator of $\frac{C\alpha^{(2)}}{aH}$ by $H$.
    However, the Peebles $C$-factor, representing the probability that a hydrogen nucleus with an electron in the $n = 2$ shell reaches the ground state before it is photoionized, is not left invariant and introduces a symmetry breaking term \cite{Peebles:1968ja}. 
    $C$ is parameterized by three key rates and is given by
    \begin{equation} \label{eq:peebles}
        C = \frac{R_{\rm{Ly\alpha}}+\Lambda_{\rm 2s-1s}}{\beta_{\rm B} + R_{\rm{Ly\alpha}}+\Lambda_{\rm 2s-1s}}.
    \end{equation}
    The escape rate of Lyman-$\alpha$ photons, $R_{\rm{Ly\alpha}}$, is
    \begin{equation} \label{eq:lyalpharate}
        R_{\rm{Ly\alpha}} = \frac{8\pi H}{\lambda^3_{\rm Ly\alpha}n_{\rm H}x_{\rm 1s}},
    \end{equation}
    where $n_{\rm H}$ is the number density of hydrogen, $x_{\rm 1s}$ is the fraction of hydrogen with an electron in the ground state, and $\lambda_{\rm Ly\alpha}$ is the Lyman-$\alpha$ wavelength given by 
    $\frac{1}{\lambda_{\rm Ly\alpha}} = \frac{3}{4}B_{\rm H}$.
    The factor of $\frac{3}{4}$ arises from the electron transitioning from the first excited state to the ground state. 
    We observe that $R_{Ly\alpha} \propto \lambda^{-3}_{\rm Ly\alpha}Hn_{\rm H}^{-1}$ which under the tFFAT scaling becomes
    \begin{equation}
        R_{\rm{Ly\alpha}} \propto \lambda^{-3}_{\rm Ly\alpha}Hn_{\rm H}^{-1} \xrightarrow{\lambda} \lambda^{-3}_{\rm Ly\alpha}(\lambda H)n_{\rm H}^{-1} \propto \lambda R_{\rm{Ly\alpha}}.
    \end{equation}
    The two-photon decay rate, $\Lambda_{\rm 2s-1s}$, scales as \cite{Spitzer1951}
    \begin{equation} \label{eq:twophotonrate}
        \Lambda_{\rm 2s-1s} \propto \alpha^8m_{\rm{e}} \xrightarrow{\lambda} (\lambda^{1/6}\alpha)^8(\lambda^{-1/3}m_{\rm{e}}) \propto \lambda \Lambda_{\rm 2s-1s}.
    \end{equation}
    The case-B photoionization rate, $\beta_{\rm B}$, is expressed as
    \begin{equation} 
        \beta_{\rm B} = \left(\frac{m_{\rm e}T}{2\pi}\right)^{3/2}e^{-B_{\rm H}/T}\alpha^{(2)}.
    \end{equation}
    Under the tFFAT scaling, $\beta_{\rm B}$ transforms as:
    \begin{equation} \label{eq:ionrate}
        \beta_{\rm B} \propto m_{\rm e}^{3/2}\alpha^{(2)}T^{3/2} \xrightarrow{\lambda} \left(\lambda^{-1/3}m_{\rm e}\right)^{3/2}\left(\lambda\alpha^{(2)}\right)T^{3/2} = \lambda^{1/2}\beta_{\rm B}.
    \end{equation}
    Employing equations \eqref{eq:lyalpharate}, \eqref{eq:twophotonrate}, and \eqref{eq:ionrate}, we scale the Peebles $C$-factor as
    \begin{equation}\label{eq:last}
        C \xrightarrow{\lambda} \frac{ R_{\rm{Ly\alpha}}+\Lambda_{\rm 2s-1s}}{\lambda^{-1/2}\beta_{\rm B} + R_{\rm{Ly\alpha}}+\Lambda_{\rm 2s-1s}}
    \end{equation}
    This introduces a minor symmetry-breaking term due to the photoionization rate $\beta_{\rm B}$ scaling as $\lambda^{-1/2}$, slightly increasing the net rate of hydrogen recombination. 
    In the following section, we numerically investigate the magnitude of symmetry breaking introduced by recombination.

\section{Phenomenological and Numerical Implementation} \label{sec:toy}

\subsection{Mimicking the MWDS in CLASS} \label{subsec:mimic}

    Here we present the mimic FFAT (mFFAT) scaling which modifies the visible baryon and photon densities according to equation \eqref{eq:scaledfriedmann}.
    The mFFAT scaling replicates the gravitational impacts introduced by a MWDS, mimics the recombination history of the tFFAT scaling, and keeps the observables protected by the FFAT scaling in check.
    While this model is schematic instead of physical (since it technically violates the COBE-FIRAS bounds), it captures all relevant physics to explore variations in $\alpha$ and $m_{\rm e}$ as credible pathways to approaching the Hubble tension.
    It also highlights the crucial role of the Thomson scattering rate in maintaining observational congruence and reaffirms the utility of FFAT scaling as a robust mechanism to safeguard cosmological observables.

    To mimic the gravitational influence that a MWDS would have on the observable Universe, we remove the COBE-FIRAS constraint from CLASS and replace dark radiation with visible radiation by increasing $T_{\rm{CMB}}$ such that $\rho_{\rm{r}} \propto T_{\rm{CMB}}^4 \xrightarrow{\lambda} \lambda^2\rho_{\rm{r}} \propto \left(\sqrt{\lambda}T_{\rm{CMB}}\right)^4.$
    However, the recombination history is exponentially sensitive to the ratio $\frac{B_{\rm H}}{T_{\rm CMB}}$.
    Thus, we must ensure that the ratio of the two remains invariant such that 
    \begin{equation} \label{eq:mimicratio}
        \frac{B_{\rm{H}}}{T_{\rm{CMB}}} \xrightarrow{\lambda} \frac{\sqrt{\lambda}B_{\rm{H}}}{\sqrt{\lambda}T_{\rm{CMB}}}.
    \end{equation}
    This maintains the ratio in the exponential of equations \eqref{eq:saha} and \eqref{eq:Xehistory}, ensuring that the changes to the recombination history remain negligible.

    We also remove the observed baryon density constraint \cite{Mossa:2020gjc} and add visible baryons as a facsimile for dark baryons.
    Scaling the visible baryon density equally scales the electron number density $n_{\rm e}$ by $\lambda^2$.
    To preserve the ratio in equation \eqref{eq:ratio}, we must choose a different scaling of $\sigma_{\rm{T}}$ as $\sigma_{\rm{T}} \xrightarrow{\lambda} \frac{1}{\lambda}\sigma_{\rm{T}}$ to counterbalance the increased $n_{\rm e}$. 
    To stay in the FFAT scaling direction, we also add additional dark matter and neutrinos, thus preserving the free streaming-radiation fraction and observables related to matter clustering.    
    The new constraints provided by equation \eqref{eq:mimicratio} and the Thomson cross section result in a different variation of fundamental constants given by
    \begin{equation}\label{eq:proxycase}
        m_{\rm{e}} \xrightarrow{\lambda} \lambda^{1/2} m_{\rm{e}} \And \alpha \xrightarrow{\lambda}  \alpha.
    \end{equation}
    As we now describe, this transformation effects recombination in a subtly different way than the tFFAT scaling, since the Peebles factor (eq.~\eqref{eq:peebles}) is serendipitously left invariant in this case.
    The escape rate of Lyman-$\alpha$ photons in the mFFAT scaling scales as
    \begin{equation}
        R_{\rm{Ly\alpha}} \propto \lambda^3_{\rm Ly\alpha}Hn_{\rm H}^{-1} \xrightarrow{\lambda} \left(\lambda^{1/2}\lambda_{\rm Ly\alpha}\right)^3(\lambda H)\left(\lambda^2n_{\rm H}\right)^{-1} \propto \lambda^{1/2} R_{\rm{Ly\alpha}}.
    \end{equation}
    The two-photon decay rate $\Lambda_{\rm 2s-1s}$ scales as
    \begin{equation}
        \Lambda_{\rm 2s-1s} \propto \alpha^8m_{\rm{e}} \xrightarrow{\lambda} \alpha^8(\lambda^{1/2}m_{\rm{e}}) \propto \lambda^{1/2} \Lambda_{\rm 2s-1s}.
    \end{equation}
    The case B photoionization rate scales as
    \begin{equation}
        \beta_{\rm B} \propto m_{\rm e}^{3/2}\alpha^{(2)}T^{3/2} \xrightarrow{\lambda} \left(\lambda^{1/2}m_{\rm e}\right)^{3/2}\left(\lambda^{-1}\alpha^{(2)}\right)\left(\lambda^{1/2}T\right)^{3/2} = \lambda^{1/2}\beta_{\rm B}.
    \end{equation}
    As a result, the mFFAT scaling leaves the Peebles factor invariant such that
    \begin{equation}
        C \xrightarrow{\lambda} \frac{\lambda^{1/2} R_{\rm{Ly\alpha}}+\lambda^{1/2} \Lambda_{\rm 2s-1s}}{\lambda^{1/2} \beta_{\rm B} + \lambda^{1/2} R_{\rm{Ly\alpha}}+\lambda^{1/2}\Lambda_{\rm 2s-1s}} = C.
    \end{equation}
    Therefore the mFFAT scaling avoids the introduction of a symmetry-breaking term associated with the photoionization rate, $\beta_{\rm B}$, as seen in the tFFAT case.
    The mFFAT scaling can thus be considered as an idealized version of the tFFAT scaling, in which the recombination symmetry-breaking effects have been suppressed.

\subsection{Testing our implementation: The ionization history} \label{sec:Xehistory}
\begin{figure}
    \centering
    \begin{subfigure}[b]{0.49\textwidth}
        \centering
        \includegraphics[width=\textwidth]{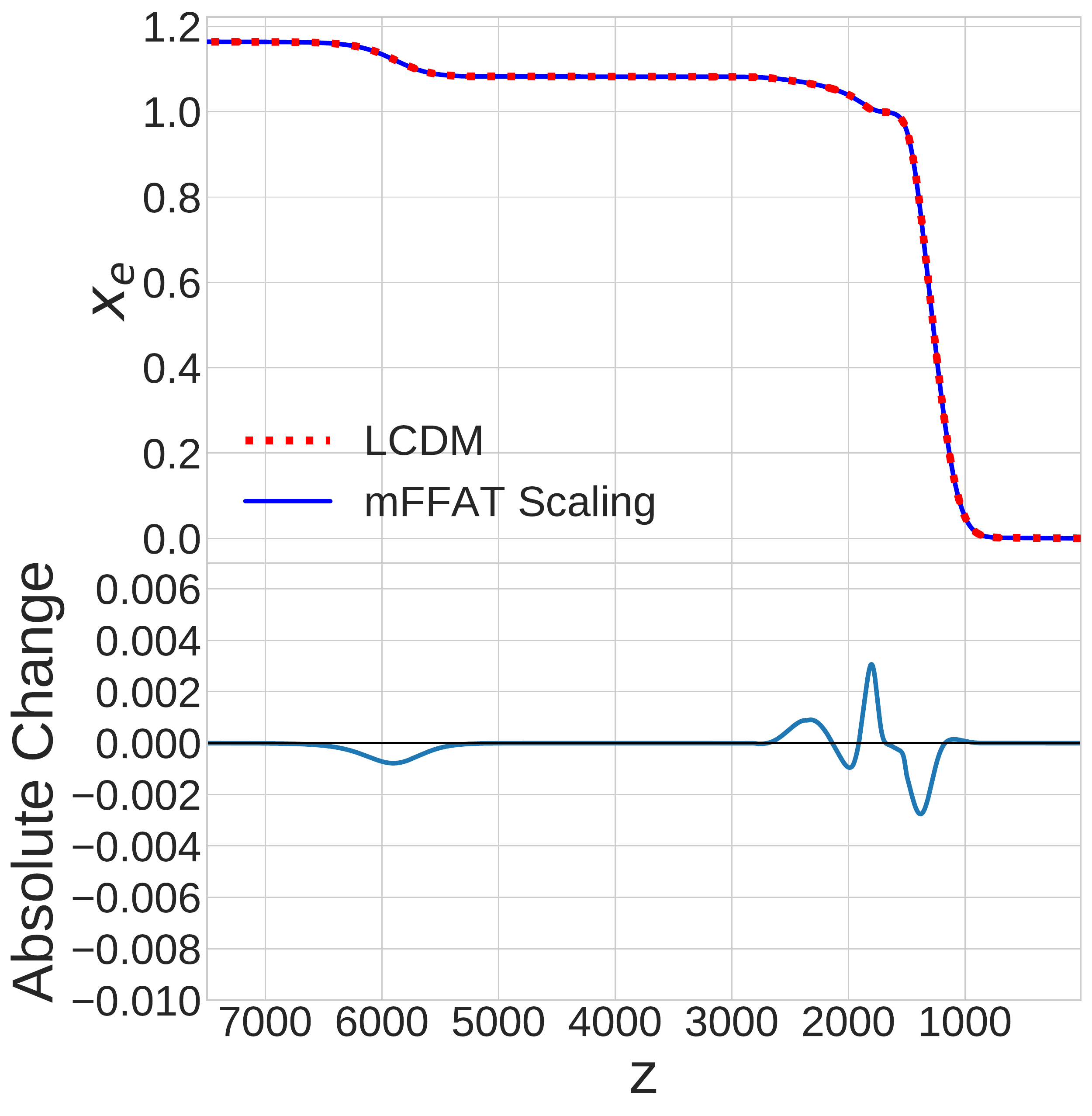}
        \caption{mFFAT scaling}
        \label{subfig:toy}
    \end{subfigure}
    \hfill
    \begin{subfigure}[b]{0.49\textwidth}
        \centering
        \includegraphics[width=\textwidth]{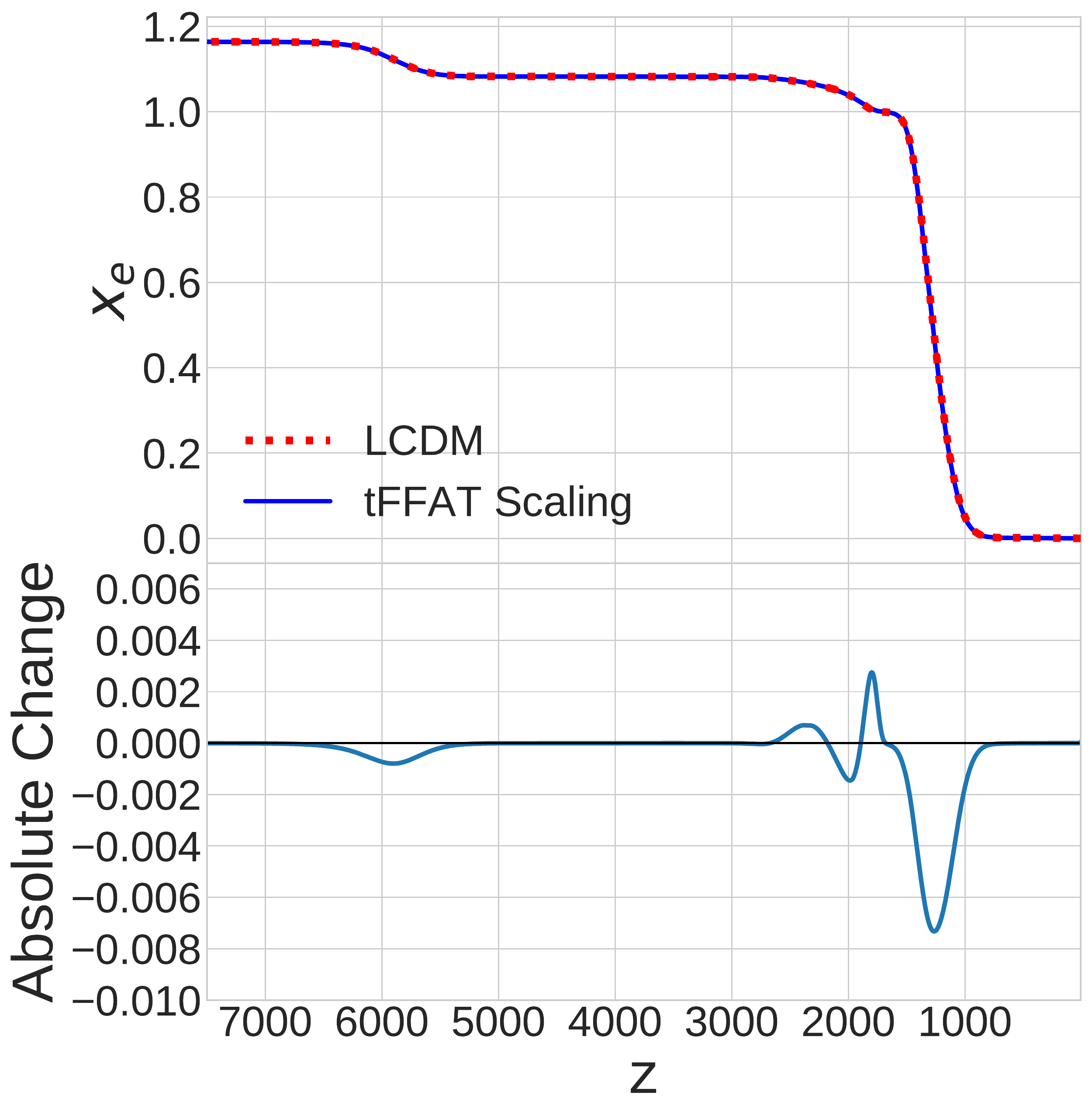}
        \caption{tFFAT scaling}
        \label{subfig:true}
    \end{subfigure}
    \centering
    \caption{Figures \ref{subfig:toy} and \ref{subfig:true} compare the expected change in the free electron fraction for the mFFAT and tFFAT models (blue) to the $\Lambda$CDM model (dotted red), scaled by $\lambda = 1.088$, corresponding to $H_0 = 73.3$ km/s/Mpc. The tFFAT $X_{\rm e}$ history in figure \ref{subfig:true} is similar to the mFFAT scaling until the onset of hydrogen recombination, where the symmetry-breaking effect from the Peebles C-factor becomes relevant.}
    \label{fig:Xehist}
    \end{figure}

\begin{figure} 
    \centering
    \begin{subfigure}[b]{0.49\textwidth}
        \centering
        \includegraphics[width=\textwidth]{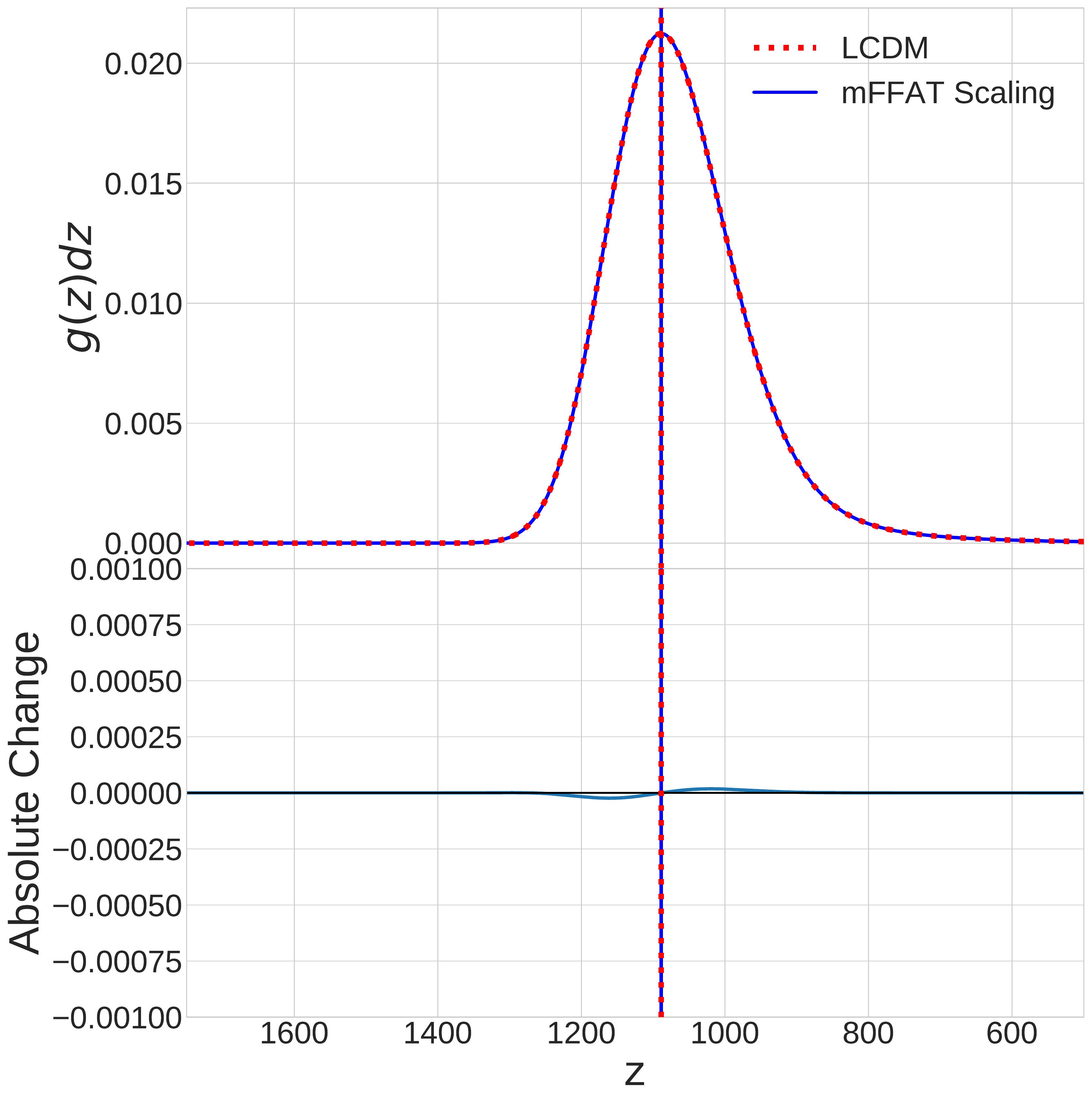}
        \caption{Mimic model with fundamental constant variation given by eq. \eqref{eq:proxycase}}
        \label{subfig:mimicvis}
    \end{subfigure}
    \hfill
    \begin{subfigure}[b]{0.49\textwidth}
        \centering
        \includegraphics[width=\textwidth]{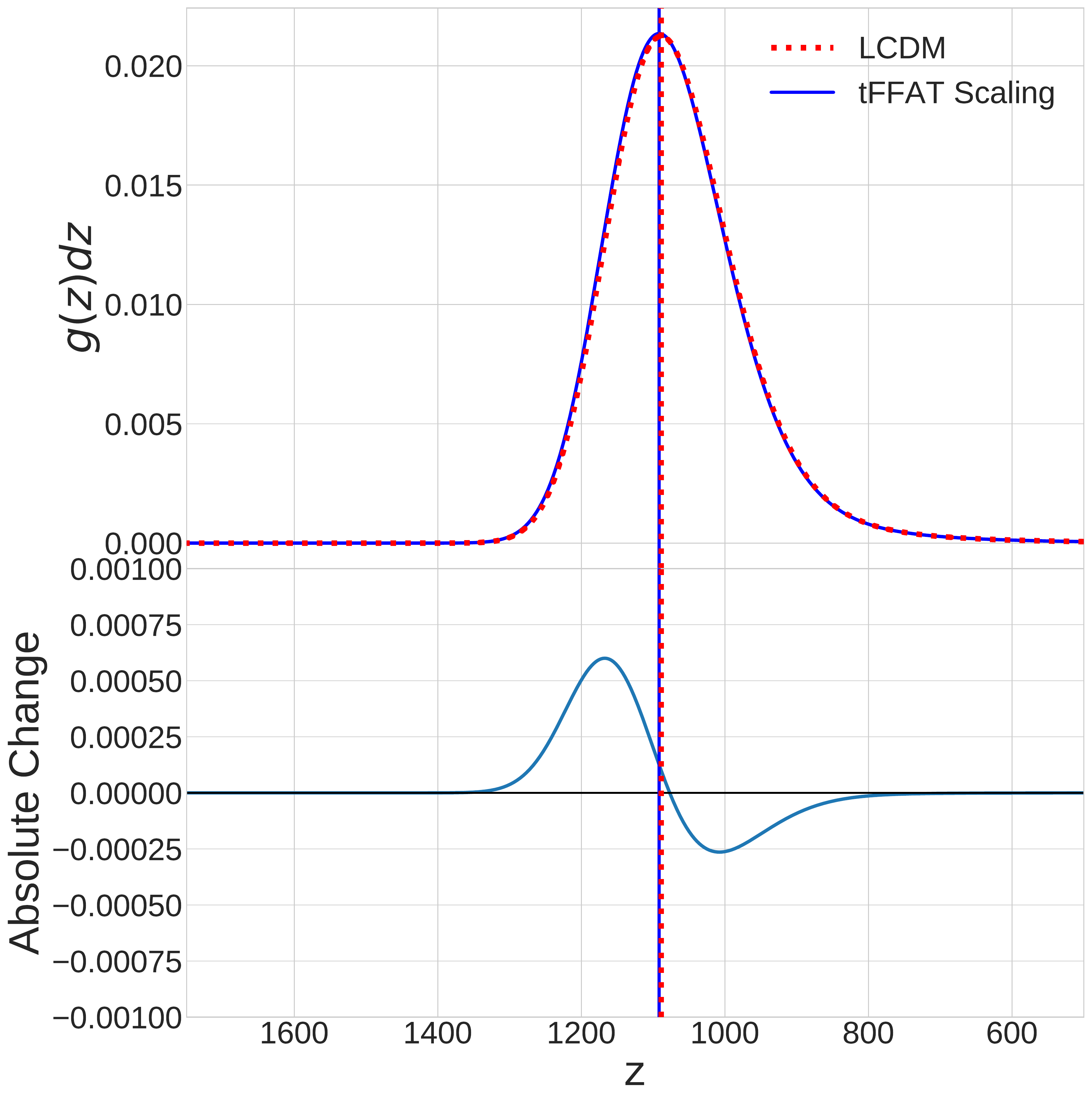}
        \caption{MWDS model with fundamental constant variation given by eq. \eqref{eq:truescale}}
        \label{subfig:truevis}
    \end{subfigure}
    \centering
    \caption{$g(z)$d$z$ is the visibility function with respect to redshift. The vertical lines indicate the location of the maximum value of the function. This figure shows that both the mFFAT and tFFAT scalings keep the visibility function's maximum and width nearly invariant compared to the $\Lambda$CDM model, even when $H_0 = 73.3$ km/s/Mpc.}
    \label{fig:visfun}
\end{figure}

   To quantify the impact of the Peebles factor's symmetry-breaking in the tFFAT scaling on the recombination history and visibility function, we use the mFFAT scaling as a baseline for comparison and perform a series of numerical tests (figures \ref{fig:Xehist} and \ref{fig:visfun}) to confirm that the mFFAT and tFFAT scalings accurately approximate the $\Lambda$CDM recombination history and visibility function.
   Figure \ref{fig:Xehist} shows $X_{\rm e}$ as a function of redshift with $\lambda = 1.088$ corresponding to $H_0 = 73.3$ km/s/Mpc for both the mFFAT and tFFAT scalings.
    The mFFAT scaling exhibits at most a 0.3\% deviation in $X_{\rm e}$ from $\Lambda$CDM due to the small symmetry-breaking term introduced from the Saha approximation, where the increased visible baryon density has shifted the equilibrium value.
    This subtle alteration is clearly seen around redshifts $z \sim 6000$ in both models\footnote{Coincidentally, the tFFAT model has the same level of symmetry-breaking for the equilibrium value, but it is due to the change in $m_{\rm e}$ and not the change in the visible baryon density.}, where the equilibrium process HeIII to HeII is very slightly changed and also influences the slight changes found at redshift $z \sim 2000$ during HeII to HeI recombination.
    While both scalings display a similar deviation from $\Lambda$CDM up until $z \sim 1500$, a significant difference arises at the onset of hydrogen recombination.\footnote{See appendix \ref{app:4level} for more details regarding the slightly different symmetry breaking introduced by the full four level model HyRec uses to compute $X_{\rm e}$.}
    This is when the recombination symmetry-breaking becomes relevant for the tFFAT model, as described at the end of section \ref{subsec:recomb}, causing at most a 0.8\% deviation.
    If the level of symmetry-breaking from the fundamental constant variation is significant enough ($\lambda$ $\gtrsim$ 1.1), it can lead to changes in CMB spectra that deviate from observations.
    However, in both cases the level of symmetry-breaking is small for values required to resolve the Hubble tension, as evidenced by the change to the visibility function seen in figure \ref{fig:visfun}.

    The visibility function represents the likelihood that a photon scatters for the last time before free-streaming away and is given by
    \begin{equation}\label{eq:gz}
        g(z) = e^{-\tau}\frac{\rm{d}\tau}{{\rm d}z}
    \end{equation}
    where $\tau$ is the optical depth to photons at redshift $z$ and is given by
    \begin{equation}\label{eq:tauz}
        \tau (z) = \int_0^z \frac{\dot{\kappa}(z)}{(1+z)H(z)} {\rm d}z.
    \end{equation}
    There are two critical features of the visibility function: the width and peak of the function.
    The peak sets the distance to the surface of last scattering, and the width corresponds to the duration of decoupling \cite{Hadzhiyska:2018mwh}.
    The mFFAT scaling remains nearly invariant for both the maximum and width of the visibility function, whereas the tFFAT scaling negligibly shifts the peak to a slightly earlier redshift.
    Though the visibility function for the tFFAT scaling is not precisely invariant due to its integration of $\dot{\kappa}$ (and thus $X_{\rm e}$) over all redshifts, the changes in the visibility function are similar to other models that increase $H_0$ while keeping CMB spectra in agreement with observation \cite{Smith:2019ihp}.

Figures \ref{fig:Xehist} and \ref{fig:visfun} also demonstrate that the free electron history does not severely constrain the variation of fundamental constants, as previously thought in ref.~\cite{Planck:2014ylh,Kaplinghat:1998ry,Hannestad:1998xp}, when combined with the FFAT scaling.
This allows for the exploration of different cosmological models with permissible constant variations that maintain a mostly invariant recombination history.
Since the mFFAT and tFFAT scalings have similar effects on recombination and maintain the similar invariance of cosmological observables, we now focus on the mFFAT scaling for computational simplicity for the remainder of this paper.
In a future publication, we will explore the tFFAT scaling in the same manner as below for the mFFAT scaling to explore the parametric freedom of the dark sector.

\subsection{Explicitly exploring the FFAT symmetry direction}

\begin{figure}[H] 
\includegraphics[width=11cm]{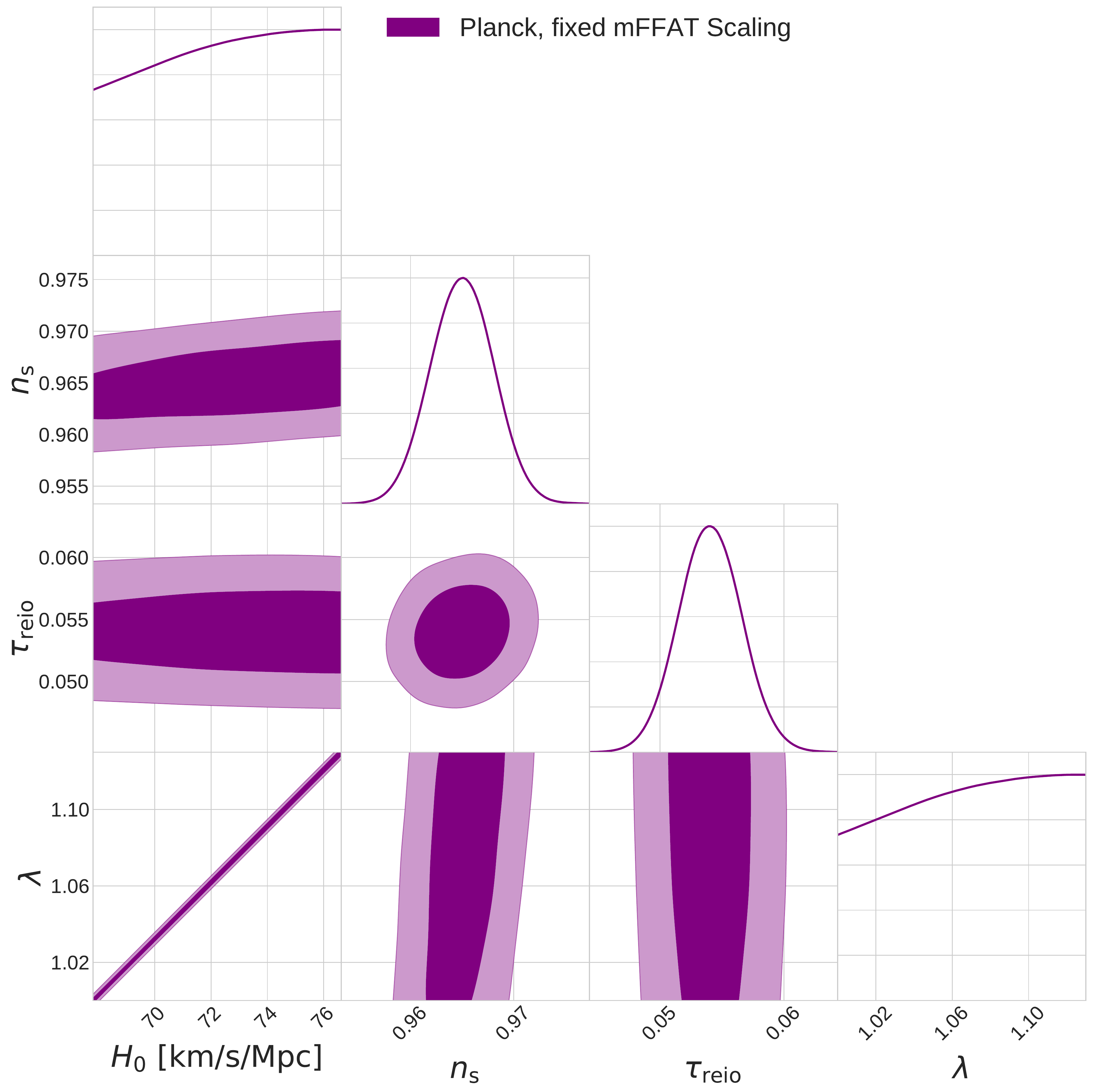}
\centering
\caption{This triangle plot shows that the mFFAT scaling allows the CMB to be made consistent with a broad range of $H_0$ values. Planck data used include TT, TE, EE, and lensing data \cite{Planck:2018vyg}. In this case, we fix the energy densities $\rho_{\rm{i}}$, $A_{\rm{s}}$, $\alpha$, $m_{\rm{e}}$, and $H$ to their $\Lambda$CDM values, and allow them to vary exclusively through the FFAT scaling parameter $\lambda$, directing MontePython to explore the symmetry direction. We have included the scaling parameter $\lambda$ to demonstrate the tight degeneracy between it and $H_0$. The upper bound for $H_0$ is determined by a limitation in the default HyRec implementation, where the interpolation tables are not expansive enough to accommodate the necessary photon energy densities.}
\label{fig:flatpost}
\end{figure}

As an initial test, we confirm numerically that the mFFAT scaling produces invariant CMB spectra while accommodating high values of $H_0$.
To do this, we fix all cosmological parameters to the typical $\Lambda$CDM values from Planck \cite{Planck:2018vyg} and introduce the variable $\lambda$ from equations \eqref{eq:ffat}.
This forces Monte Python \cite{Brinckmann:2018cvx}, the Monte-Carlo-Markov-chain sampler we use to compare the mFFAT model to data, to explicitly explore the FFAT symmetry direction.
As expected, figure \ref{fig:flatpost} shows that this `fixed' mFFAT scaling invariantly scales $H_0$, flattening it's one-dimensional posterior.
Only $H_0$, $n_{\rm{s}}$, and $\tau_{\rm{reio}}$ are displayed, while traditional parameters such as $\omega_{\rm{b}}$, $\omega_{\rm{cdm}}$, and $A_{\rm s}$ are scaled from their $\Lambda$CDM values according to equations \eqref{eq:ffat}.
The flattening of the posterior demonstrates the effectiveness of the mFFAT scaling in invariantly increasing $H_0$, while maintaining consistency with observed CMB spectra.

\section{Results} \label{sec:results}

\subsection{Naturally exploring the FFAT scaling}
\begin{figure}[H] 
\includegraphics[width=15cm]{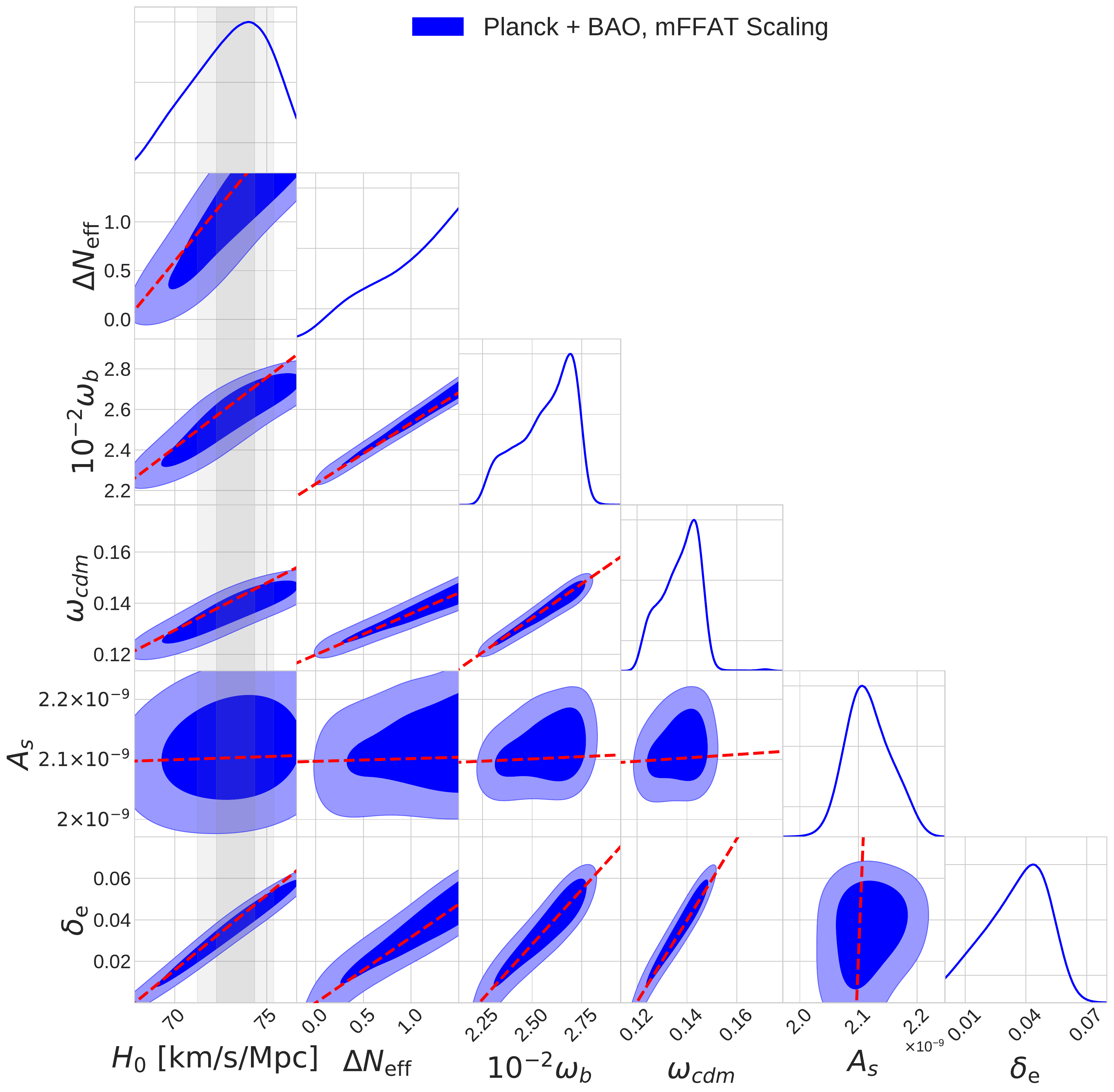}
\centering
\caption{This triangle plot demonstrates the natural realization of the FFAT scaling. The red dashed lines indicate the direction of the FFAT symmetry. Note that the contours agree with the red lines within 1$\sigma$. We see a 1$\sigma$ range for the posterior on $H_0$ that extends well beyond the typical $\Lambda$CDM model. A high value of $H_0$ is preferred with no data representing the local measurement. The grey band represents the local measurement of $H_0 = 73.3 \pm 1.04$ km/s/Mpc from ref.~\cite{Riess:2021jrx}.}
\label{fig:naturalevo}
\end{figure}

In our second test of the mFFAT scaling, we allow all cosmological parameters to vary and remove the explicit enforcement of the FFAT scaling.
As illustrated in Figure \ref{fig:naturalevo}, the FFAT scaling direction, indicated by the dashed red lines, naturally emerges form the MCMC exploration.
By allowing all cosmological parameters to vary, we find that the MCMC sampler naturally broadens the posterior on $H_0$, enhancing our understanding of the relationship between the Thomson scattering rate and the Hubble tension.
We find the correct scaling between $N_{\rm{eff}}$, $\omega_{\rm{cdm}}$, and $\omega_{\rm{b}}$, in accordance with equation \eqref{eq:scaledfriedmann}. 
We also define the variation of $m_{\rm e}$ by $\delta_{\rm e} = \frac{\tilde{m}_{\rm e}}{m_{\rm e}} - 1$ where $\tilde{m}_{\rm e}$ is the varied electron mass.
For a list of priors used to generate figures \ref{fig:flatpost} and \ref{fig:naturalevo}, please see table \ref{tab:priors}.

We note that as $\delta_{\rm e}$ approaches 5\%, corresponding to a value of $\lambda \sim 1.1$ ($H_0$ = 74.4 km/s/Mpc), the ionization balance shifts, and the symmetry breaking effects introduced by recombination cause a deviation from the FFAT symmetry direction.
This is reflected in the one-dimensional posterior for $\delta_{\rm e}$, which finds very high values of $\delta_{\rm e}$ to be increasingly less likely.
$\Delta N_{\rm{eff}}$ shown in figure \ref{fig:naturalevo} is a derived parameter given by
\begin{equation}
    \Delta N_{\rm{eff}} = \left(\frac{\Tilde{T}^4}{T_{\rm{COBE}}^4} - 1\right) \left(3.044 + \frac{1}{\frac{7}{8}\left(\frac{4}{11}\right)^{4/3}} \right)
\end{equation}
where $\Tilde{T}$ is the CMB temperature that is allowed to vary by the MCMC representing the additional radiation density, and $T_{\rm{COBE}}$ is the COBE-FIRAS temperature measurement of the CMB equal to 2.7255 K \cite{Fixsen:2009ug}.
The prior volume is limited for $\Delta N_{\rm eff}$ from a limitation of default HyRec interpolation tables for the CMB temperature.

When comparing the best-fit parameters from figure \ref{fig:naturalevo} to the best-fit $\Lambda$CDM values, it becomes evident that the mFFAT scaling prefers a higher value of $H_0$ and is able to fit Planck and BAO data as effectively as the $\Lambda$CDM model, with a corresponding $\Delta\chi^2 = -2.3$.
This improvement of $\chi^2$ is expected as we have included two additional parameters into the model.
However, the prior volume effect listed above may influence the shape of posterior, and a larger volume may find a preference for even higher values of $H_0$.
Taken together, these findings indicate that fundamental constant variation with the FFAT scaling offers a promising approach for investigating the Hubble tension.

\begin{figure}[]
\centering
    \begin{subfigure}[b]{0.49\textwidth}
        \centering
        \includegraphics[width=\textwidth]{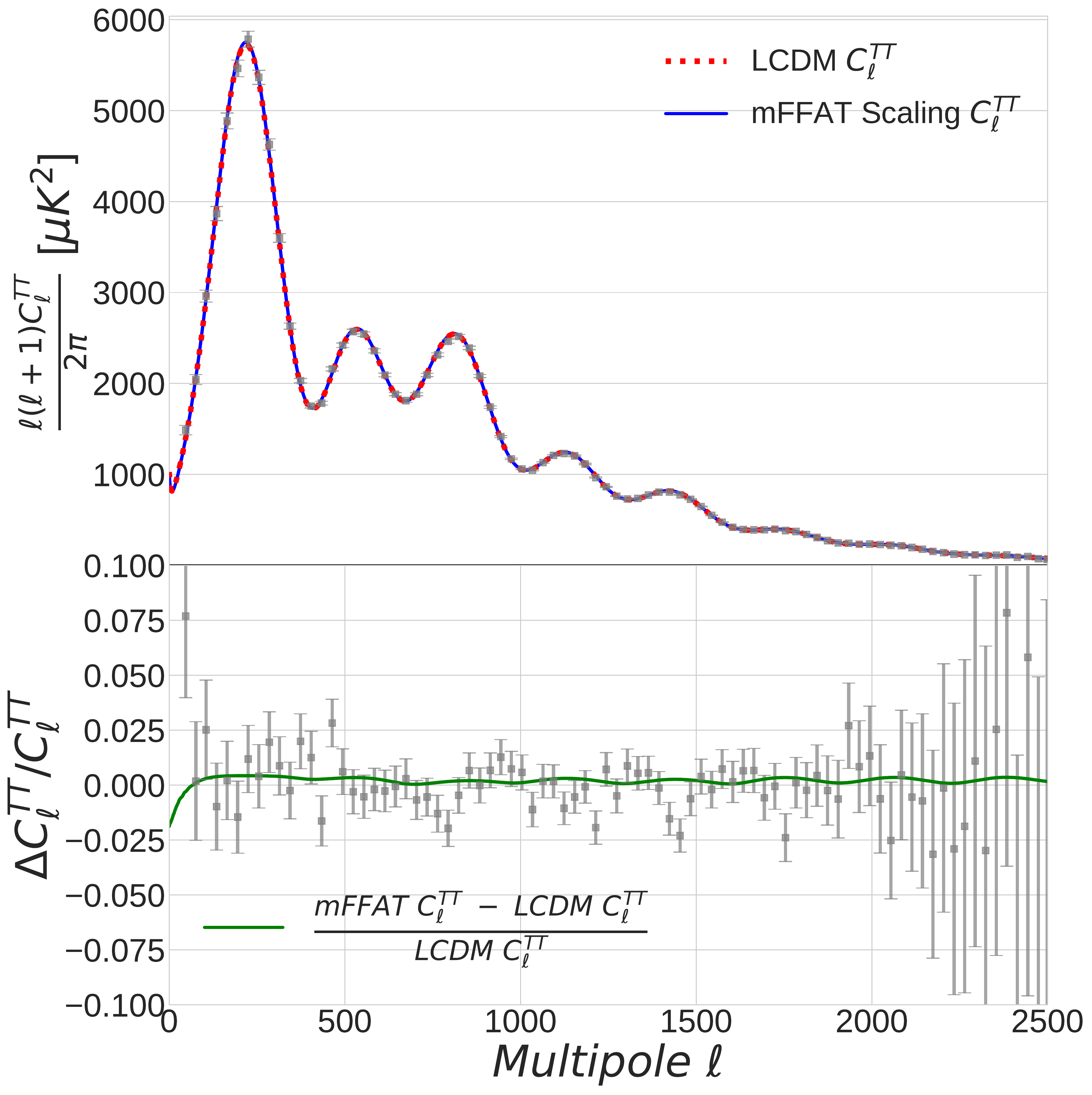}
        \caption{CMB Temperature Spectra}
        \label{subfig:TTspectra}
    \end{subfigure}
    \hfill
        \begin{subfigure}[b]{0.49\textwidth}
        \centering
        \includegraphics[width=\textwidth]{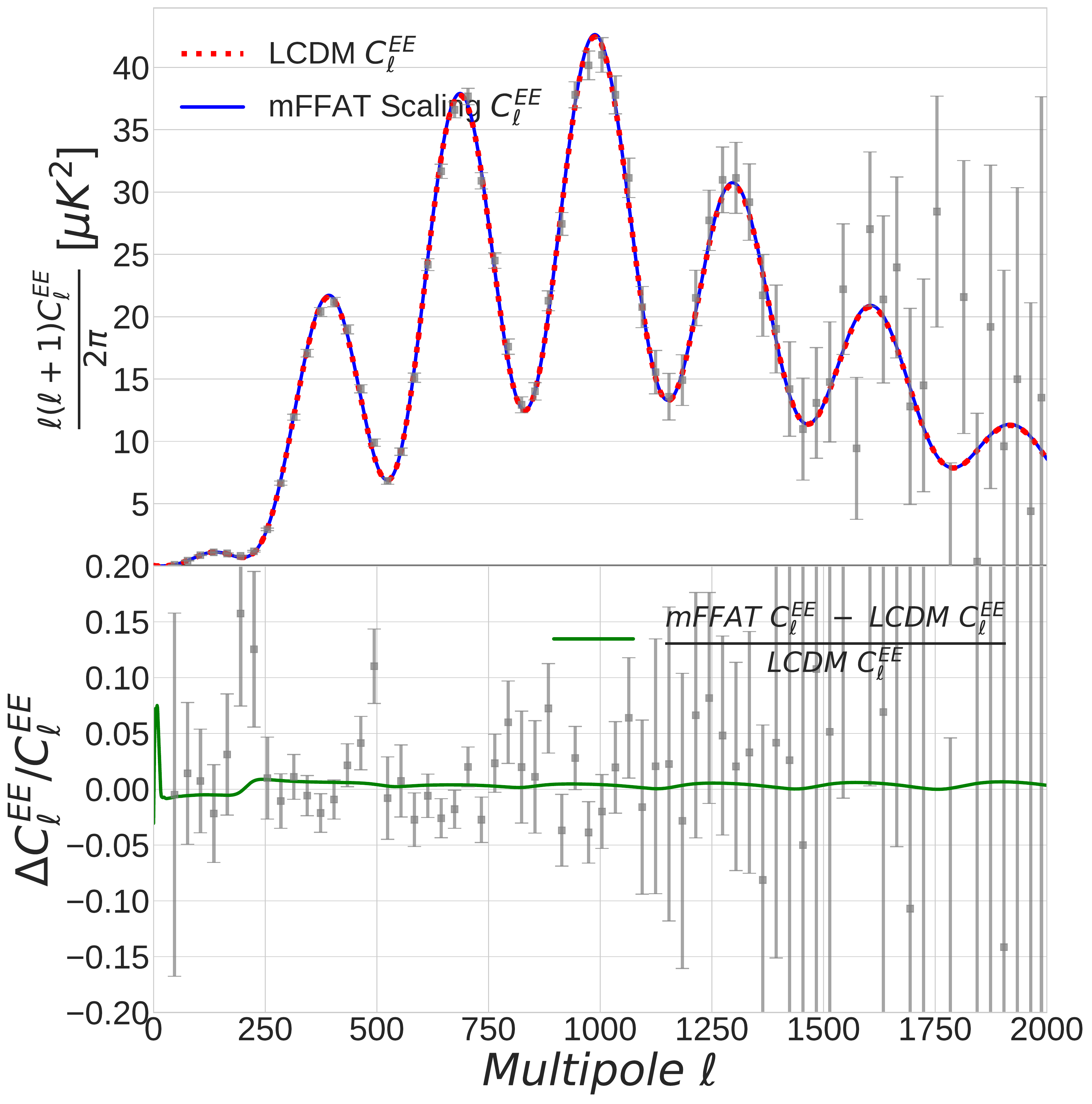}
        \caption{CMB Polarization Spectra}
        \label{subfig:EEspectra}
    \end{subfigure}
\centering
\caption{These plots compare the TT and EE CMB spectra for the $\Lambda$CDM model and the mFFAT scaling utilizing the best fit values from figure \ref{fig:naturalevo} ($H_0$ = 73.97 km/s/Mpc). The grey points represent a subsample of measured values from Planck, accompanied by their respective error bars. The mFFAT scaling demonstrates remarkable agreement with Planck observations in both the TT and EE spectra, while preserving locally measured $H_0$ values.}
\label{fig:spectra}
\end{figure}

The CMB temperature-temperature (TT) and polarization-polarization (EE) spectra obtained with the mFFAT scaling best fit values from figure \ref{fig:naturalevo} are compared to those of a $\Lambda$CDM model in Figure \ref{fig:spectra}.
The CMB spectra exhibit a high degree of invariance, concurrently accommodating both cosmological and local observations of the expansion rate, aligning with the predictions made in Section \ref{subsec:amp}.
This example serves as a successful demonstration of how the FFAT scaling can be utilized to explore a wider range of cosmological models, without sacrificing agreement with observational data.

\subsection{Relationship with other work}
The interplay between variations in fundamental constants and the Hubble tension has been explored in previous works, though in different scenarios \cite{Sekiguchi:2020teg, Martins:2017yxk, Hart:2017ndk, Hart:2019dxi, Hart:2021kad,Hart:2022agu,Hart:2020,Lee:2022gzh, Schoneberg:2021qvd, Webb:1998cq, Murphy:2000pz, Sandvik:2001rv, Uzan:2002vq, Webb:2010hc, Srianand:2004mq, Chand:2004ct, Mota:2003tc, King:2012id, Fischer:2004jt, Dixit:1987at, Mota:2003tm, Coc:2006sx, Huntemann:2014dya, Cowie:1995sz, Petrov:2005pu, Reinhold:2006zn, Calmet:2002ja, Uzan:2004qr, Fritzsch:2016ewd, Flambaum:2007my, Berengut:2010yu, Calmet:2006sc, Chiba:2006xx, Karshenboim:2003jc, Li:2021dvt, Bekenstein:2002wz}.
For example, reference \cite{Sekiguchi:2020teg} investigated the effect of varying the electron mass to allow for an earlier recombination time.
They outlined four criteria for a "successful" cosmological solution to the Hubble tension.
These criteria include: maintaining the CMB power spectra similar to $\Lambda$CDM, reducing sound horizon size by $\sim 10\%$, ensuring the angular size of the sound horizon aligns with $\Lambda$CDM, and and not worsening the fit to BAO, SNeIa, and other low-redshift distance measurements.
Notably, these four criteria are also naturally satisfied along the FFAT scaling direction.
However, a deviation from the FFAT scaling occurs in their model when increasing the physical matter density, which alters the redshift of matter-dark energy equality and limits their model in increasing $H_0$. Furthermore, the variation of $m_e$ proposed in reference \cite{Sekiguchi:2020teg} is inverse to what the FFAT scaling suggests, aiming for a larger electron mass. These distinctions could illuminate why models on the FFAT scaling direction demonstrate an increase in $H_0$ compared to the model presented by reference \cite{Sekiguchi:2020teg}, where only $m_e$ is varied.

\begin{figure} 
\includegraphics[width=10cm]{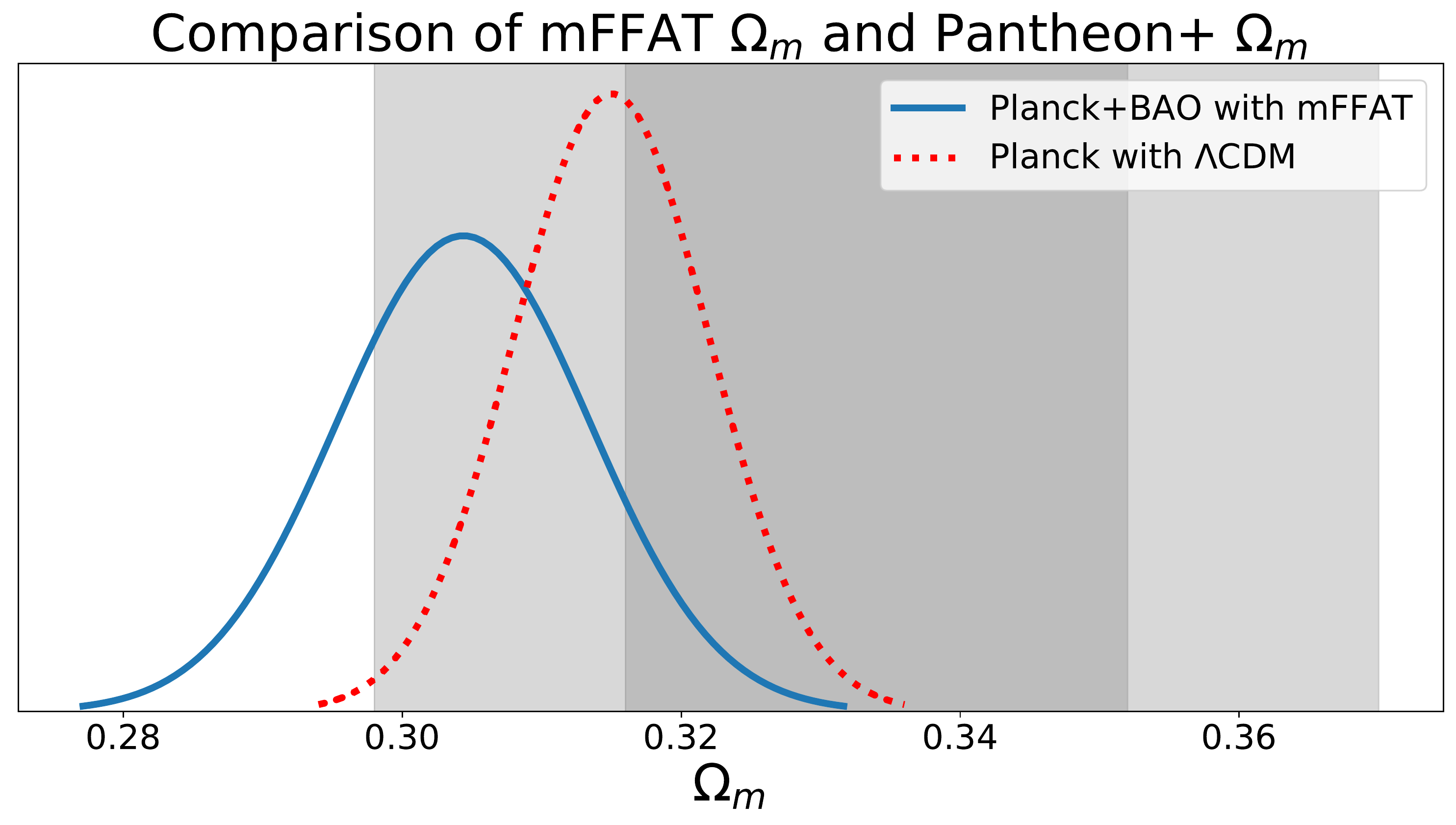}
\centering
\caption{Posteriors of $\Omega_{\rm m}$ shown up to 3$\sigma$ for the mFFAT model with Planck+BAO data and the baseline $\Lambda$CDM model with Planck data. In grey is the Pantheon+ derived value of $\Omega_m$ representing the 1$\sigma$ (dark grey) and 2$\sigma$ (light grey) regimes.}
\label{fig:Omega_m}
\end{figure}

In a different approach, reference \cite{Lee:2022gzh} utilizes Fisher-bias information to identify data-driven extensions to $\Lambda$CDM involving variation of fundamental constants.
Their findings align with the notion that a modified recombination history could ease the Hubble tension when solely considering CMB data.
They find that when, for example, the electron mass is allowed to vary it is accompanied by a decrease in the physical baryon and cold dark matter densities.
However, much like reference \cite{Sekiguchi:2020teg}, their methodology encounters similar limitations: the change in the physical matter density changes the redshift of matter-dark energy equality, veering the models away from the FFAT scaling direction.
Indeed, because the FFAT scaling direction ensures that ratio of the observed matter density to the critical density is left consistent compared to $\Lambda$CDM (see sec. \ref{subsec:ff}), it is able to accommodate values of $\Omega_{\rm m}$ which are less than 2$\sigma$ discrepant with recent supernova surveys like Pantheon+ \cite{Brout:2022vxf}, as shown in figure \ref{fig:Omega_m}, avoiding the constraints prior models encountered. 

\section{Conclusion} \label{sec:conc}
    In this research, we have furthered the understanding of the Hubble tension through the employment of the Free-Fall, Amplitude, and Thomson (FFAT) scaling in conjunction with a variation of fundamental constants given by equation \eqref{eq:truescale}.
    This alternative approach substantiates the findings of previous studies \cite{Cyr-Racine:2021oal,Ge:2022qws}, demonstrating that the FFAT scaling successfully can bring the local and distant measurements of $H_0$ into concordance with each other.
    Here, we briefly outline the key results of our study:
    \begin{itemize}
    \item The FFAT scaling with fundamental constant variation successfully achieves high values of $H_0$, corroborating the results of refs.~\cite{Cyr-Racine:2021oal,Ge:2022qws} through an alternative approach and emphasizes the critical role the Thomson scattering rate takes in the context of the Hubble tension.
    \item By safeguarding the ratio of the Thomson scattering rate to the background expansion rate, the FFAT scaling combined with the variation of fundamental constants significantly mitigates the restrictions on fundamental constant variation during the epoch of recombination.
    This protection of the recombination history allows for greater flexibility and exploration in cosmological models which vary fundamental constants.
    \item An MCMC exploration of the FFAT scaling broadens the posterior for $H_0$, reconciling the local and distant measurements of $H_0$ utilizing only LSS and CMB data. Including a prior on the local measurement of $H_0$ would result in a significant improvement to the fit compared to $\Lambda$CDM.
    \end{itemize}
    The methodology proposed in this research lays a robust groundwork for subsequent investigations into the Hubble tension and physics beyond the standard models of particle physics and cosmology.
    Our study highlights three promising avenues for future exploration.
    Firstly, the incorporation of the complete MWDS into CLASS will allow for exploration of the dark sector parameter space beyond simple energy densities.
    Secondly, a thorough examination of the relationship between the FFAT scaling and BBN may broaden the tight constraints set on the early Universe by primordial element abundances.
    Lastly, the development of particle physics models which encapsulates the MWDS and fundamental constant variation can potentially elucidate the enigmatic nature of the dark sector.

\acknowledgments
This work was supported by the National Science Foundation (NSF) under grant AST-2008696. F.-Y. C.-R. would like to thank the Robert E.~Young Origins of the Universe Chair fund for its generous support. We would like to thank the UNM Center for Advanced Research Computing, supported in part by the NSF, for providing the research computing resources used in this work. 

\begin{appendices}
\section{Exact scaling transformations for the generalized Peeble's C-factor}
\label{app:4level}
The following discussion is heavily based on the original HyRec publications.
For more details and a clear discussion of the recombination process, please see refs.~\cite{Lee:2020obi,Ali-Haimoud:2010hou} and the references therein.
The generalized Peeble's C-factor \cite{Lee:2020obi,Ali-Haimoud:2010hou} used by HyRec-2 to compute the free electron fraction can be given by
\begin{equation}
    C_{\rm 2\ell} = \frac{\mathcal{R_{\rm 2\ell,1s}}+\mathcal{R_{\rm 2\ell,2\ell^{\prime}}}\frac{\mathcal{R_{\rm 2\ell^{\prime},1s}}}{\Gamma_{\rm 2\ell^{\prime}}}}{\mathcal{B_{\rm 2\ell}} + \mathcal{R_{\rm 2\ell,2\ell^{\prime}}}\frac{\mathcal{B_{\rm 2\ell^{\prime}}}}{\Gamma_{\rm 2\ell^{\prime}}}+\mathcal{R_{\rm 2\ell,1s}}+\mathcal{R_{\rm 2\ell,2\ell^{\prime}}}\frac{\mathcal{R_{\rm 2\ell^{\prime},1s}}}{\Gamma_{\rm 2\ell^{\prime}}}}
\end{equation}
where $\ell$ = p if $\ell^{\prime}$ = s, $\ell$ = s if $\ell^{\prime}$ = p, and $\Gamma_{\rm 2\ell}$ is given by
\begin{equation}
    \Gamma_{\rm 2\ell}= \mathcal{B_{\rm 2\ell}} + \mathcal{R_{\rm 2\ell,2\ell^{\prime}}}+\mathcal{R_{\rm 2\ell,1s}}.
\end{equation}
$\mathcal{B_{\rm 2\ell}}$ is the effective photoionization rate of the 2$\ell$ level, $\mathcal{R_{\rm i,j}}$ is the transition rate from the $i^{th}$ level to the $j^{th}$ level, and $\Gamma_{\rm 2\ell}$ is the effective inverse lifetime of the $2\ell$ level.
This generalized form represents the probability that an atom in the $2\ell$ state reaches the ground state by transitioning to the $n = 2$ state first or by direct transitions. 
Before we analyze the generalized C-factors, we must make the following assumptions that are used by HyREC; first, the rate of transitions from the 2s to 1s state are given by $\Lambda_{\rm 2s,1s}$.
Second, the rate of transitions from the 2p to 1s state are given by $R_{\rm Ly\alpha}$.
Therefore, we can write the $C_{\rm 2s}$ and $C_{\rm 2p}$ as
\begin{equation}
    C_{\rm 2s} = \frac{\Lambda_{\rm 2s,1s} + \mathcal{R_{\rm 2s\rightarrow 2p}}\frac{R_{\rm Ly\alpha}}{\Gamma_{\rm 2p}}}{\Gamma_{\rm 2s} - \mathcal{R_{\rm 2s\rightarrow 2p}}\frac{\mathcal{R_{\rm 2s\rightarrow 2p}}}{\Gamma_{\rm 2p}}}
\end{equation}
\begin{equation}
    C_{\rm 2p} = \frac{R_{\rm Ly\alpha} + \mathcal{R_{\rm 2s\rightarrow 2p}}\frac{\Lambda_{\rm 2s,1s}}{\Gamma_{\rm 2s}}}{\Gamma_{\rm 2p} - \mathcal{R_{\rm 2s\rightarrow 2p}}\frac{\mathcal{R_{\rm 2s\rightarrow 2p}}}{\Gamma_{\rm 2s}}}
\end{equation}
The transition rate from the 2s to 2p state is proportional to the fundamental constants such that
\begin{equation}
    \mathcal{R_{\rm 2s\rightarrow 2p}} \propto \alpha^5 m_{\rm e}
\end{equation}
We see that for both the mFFAT and tFFAT variations, $\mathcal{R_{\rm 2s\rightarrow 2p}}$ scales as $\lambda^{1/2}$.
\subsection{tFFAT scaling}
The tFFAT scaling retains the symmetry breaking presented above, primarily due to $\Lambda_{\rm 2s,1s}$ and $R_{\rm Ly\alpha}$ scaling as $\lambda$ rather than $\lambda^{1/2}$.
This causes the effective inverse lifetimes of the $2\ell$ states, $\Gamma_{\rm 2\ell}$ to not be invariant under the transformation such that
\begin{equation}
    \Gamma_{\rm 2\ell} \xrightarrow{\lambda} \lambda^{1/2}(\mathcal{B_{\rm 2\ell}} + \mathcal{R_{\rm 2\ell,2\ell^{\prime}}}+ (\lambda^{1/2})\mathcal{R_{\rm 2\ell,1s}}),
\end{equation}
thus, reducing the effective lifetime of the $2\ell$ level, corresponding to an increase in the rate of overall recombination.
This agrees with our conclusions in section \ref{subsec:recomb}.
\subsection{mFFAT scaling}
Similar to our discussion in section \ref{subsec:mimic}, the mFFAT scaling leaves the generalized C-factors invariant.
All rates that enter the system ($\mathcal{B_{\rm 2\ell}}$, $R_{\rm Ly\alpha}$, $\mathcal{R_{\rm 2s\rightarrow 2p}}$, $\Lambda_{\rm 2s,1s}$, $\Gamma_{\rm 2s}$, and $\Gamma_{\rm 2s}$) all scale as $\lambda^{1/2}$.
This leaves $C_{\rm 2 \ell}$ also invariant under the mFFAT scaling.

\section{Prior bounds for MCMC exploration}
\begin{table}[H]
\begin{center}
\begin{tabular}{|| c | c | c ||}
  \hline
  & Figure \ref{fig:flatpost}      & Figure \ref{fig:naturalevo} \\ 
  \hline
 $\omega_{\rm b}$   & 0.02238*       & [0.02, 0.0322] \\
 \hline
 $\omega_{\rm cdm}$ & 0.12011*       & [0.1, 0.173] \\
 \hline
 $h$                & 0.6781*        & [0.6, 0.75]  \\
 \hline
 $A_{\rm s}$        & 2.101e-9*      & [None, None]  \\
 \hline
 $n_{\rm s}$              & [None, None]  & [None, None]  \\
 \hline
 $\tau_{\rm reio}$      & [0.004, None] & [0.004, None]  \\
 \hline
 $T_{\rm CMB}$      & 2.7255*        & [2.7, 2.87]  \\
 \hline
 $\delta_{\rm e}+1$ & 1.0*           & [1.0, 1.095] \\
 \hline
 $\lambda$          & [1.0, 1.13]   & -  \\
 \hline
\end{tabular}
\caption{This table enumerates the prior bounds used to generate the contour plots of figures \ref{fig:flatpost} and \ref{fig:naturalevo}. A * indicates the reference values ($\lambda=1$) of the parameters that are then scaled according to eq.~\eqref{eq:ffat} as $\lambda$ is varied in the MCMC chains. }
\label{tab:priors}
\end{center}
\end{table}

\end{appendices}

\bibliographystyle{JHEP.bst}
\bibliography{main}

\end{document}